\documentclass{elsarticle}
\usepackage{hyperref}
\usepackage{multirow}%
\usepackage{graphicx}
\usepackage{natbib}
\usepackage{textcomp}
\usepackage{xcolor}

\usepackage{url}
\usepackage{amstext}
\usepackage{amssymb}
\usepackage{amsmath}
\usepackage{algorithm}
\usepackage{algorithmic}
\usepackage{multicol}
\usepackage{tikz}
\usepackage{lettrine}
\allowdisplaybreaks
\usepackage[export]{adjustbox}
\usepackage{ragged2e}
\usepackage{float}
\usepackage{url}
\usepackage{tabularx}
\usepackage{booktabs}
\usepackage{array}
\usepackage{pgfplots}
\usepackage{subfigure}
\usepackage{amsthm}
{}
{}
{}

\begin{document}

\begin{frontmatter}

\title{FedMUP: Federated Learning driven Malicious User Prediction Model for Secure Data Distribution in Cloud Environments}

\author[mymainaddress]{Kishu Gupta}
\author[mysecondaryaddress,mysixthaddress]{Deepika Saxena \corref{mycorrespondingauthor}}
\cortext[mycorrespondingauthor]{Corresponding author}
\ead {13deepikasaxena@gmail.com, deepika@u-aizu.ac.jp}
\author[mythirdaddress]{Rishabh Gupta}
\author[myfourthaddress]{Jatinder Kumar}
\author[myfifthaddress,mysixthaddress]{Ashutosh Kumar Singh} 

\address[mymainaddress]{Department of Computer Science and Engineering, National Sun Yat-sen University, Kaohsiung, 80424, Taiwan}
\address[mysecondaryaddress]{Department of Computer Science \& Engineering, University of Aizu, Aizuwakamatsu, Fukushima, Japan}
\address[mythirdaddress]{Department of Computer Applications, SRM Institute of Science and Technology, NCR Campus-Ghaziabad, 201204, India}
\address[myfourthaddress]{Department of Computer Applications, National Institute of Technology, Kurukshetra, 136119, India}
\address[myfifthaddress]{Department of Computer Science and Engineering, Indian Institute of Information Technology, Bhopal, 462003, India}
\address[mysixthaddress]{Department of Computer Science, The University of Economics and Human Sciences, Warsaw, 01043, Poland}

\begin{abstract}
Cloud computing is flourishing at a rapid pace. Significant consequences related to data security appear as a malicious user may get unauthorized access to sensitive data which may be misused, further. This raises an alarm-ringing situation to tackle the crucial issue related to data security and proactive malicious user prediction. This article proposes a \textbf{Fed}erated learning driven \textbf{M}alicious \textbf{U}ser \textbf{P}rediction Model for Secure Data Distribution in Cloud Environments (\textbf{FedMUP}). This approach firstly analyses user behavior to acquire multiple security risk parameters. Afterward, it employs the federated learning-driven malicious user prediction approach to reveal doubtful users, proactively. FedMUP trains the local model on their local dataset and transfers computed values rather than actual raw data to obtain an updated global model based on averaging various local versions. This updated model is shared repeatedly at regular intervals with the user for retraining to acquire a better, and more efficient model capable of predicting malicious users more precisely. Extensive experimental work and comparison of the proposed model with state-of-the-art approaches demonstrate the efficiency of the proposed work. Significant improvement is observed in the key performance indicators such as malicious user prediction accuracy, precision, recall, and f1-score up to 14.32\%, 17.88\%, 14.32\%, and 18.35\%, respectively. 
\end{abstract}


\end{frontmatter}

\begin{center}
	\footnotesize{\textbf{This article has been accepted for publication in Applied Soft Computing Journal 1568-4946/© 2024 Elsevier B.V. Published by Elsevier Ltd. All rights reserved. \\Citation information: DOI 10.1016/j.asoc.2024.111519.\\ Received 29 May 2023; Received in revised form 15 January 2024; Accepted 14 March 2024; Available online 20 March 2024. \\Personal use of this material is permitted. Permission from Elsevier must be obtained for all other uses, in any current or future media, including reprinting/republishing this material for advertising or promotional purposes, creating new collective works, for resale or redistribution to servers or lists, or reuse of any copyrighted component of this work in other works. This work is freely available for survey and citation.}}
\end{center}
%
%


\section{Introduction}\label{secint}
Data is the key to growth and is evolving at a tremendous pace. Data sharing over cloud platforms has emerged as a new fundamental need of any organization, aspiring to excel \cite{song2022public, saxena2022high}. Considerable facilities to accommodate data at a reasonable cost \cite{wei2016secure, singh2021quantum}, data analysis with enormous computation power, and data sharing among various stakeholders for further utilization of data \cite{gupta2022iot, saxena2021secure, icdam2021} put forwards the huge aspects of the cloud assistance. At present almost 94\% of organizations employ cloud services \cite{cloud} eventually leading to data contributors' losing control over data \cite{gupta2022differential, chandra2021}. Moreover, despite multiple amazing facilities, cloud service may be abused for data compromise thus pushing sensitive data at high risk \cite{shen2018enabling, li2019meta, deps1-10243065}. This data vulnerability might make data contributors hesitant to outsource data to the cloud platform \cite{yin2022efficient, rgupta_singh2022, deps2-10272307}. As per the 'Cost of a Data Breach Report 2022' \cite{IBM} by 'IBM Security' almost 83\% of organizations considered have encountered more than one data breach, and more interestingly 45\% data breaches occurred over the cloud platform. Another, 'Data Breach Investigations Report (DBIR)' \cite{DBIR} stipulates that data breach incidence is swollen up by 2.6\% out of which 82\% of breaches involved the human element classifying more than 80\% breaches induced externally, approximately 18\% breaches emanated internally and reaming due to other factors. These findings highlight the severity of data protection and hence ultimately served as the motivation behind this research to formulate a method capable enough to tackle these data security issues, proactively.
\par Traditionally, a) watermarking, and b) probability-based approaches are extensively described and utilized to determine the possible malicious user \cite{martinsurvey2021}. Watermarking mandates some sort of data alterations into actual data to detect malicious ones later through the retrieval of embedded data. But this embedded data itself might experience alterations \cite{shehab2007watermarking, saxena2021osc, almehmadi2022novel}. On the other hand, probability-oriented approaches deploy the computational efficacy of machine learning and perform well but are not practical all the time due to the extensive need for data sharing which again itself raises data security concerns \cite{gupta2020mlpam, rgupta2022}. Moreover, the current techniques determine the malicious user after the occurrence of a data breach. Therefore, there is a critical requirement to develop an approach that can impart secure data storage, sharing, and proactive malicious user detection considering performance up-gradation. Federated learning-driven data security approaches appear as an ultimate, up-to-mark, and most promising solution for such an issue \cite{GONG2023, MOTHUKURI2021619}.
\par Considering the aforementioned issues and challenges a novel Federated Learning driven Malicious User Prediction approach for data security is proposed in this article. The model comprises two units a) Analyze the user behavior, and b) Malicious user prediction. Various security parameters obtained from user behavior analysis based on available current and historical details are passed to the Federated learning (FedL) Machine Learning-based Malicious User Prediction unit for further analysis. In the process, a central global model is obtained from the aggregation of the local models, while local nodes operate local training at their devices with their local data, without actually sharing this crucial data. Therefore, this prediction model intensifies data security by confining data communications needs and proactive detection of a user as 'malicious' or 'non-malicious'. To the best of the author's knowledge, the FedMUP is the first model to predict malicious users proactively using a federated learning environment for controlling and mitigating cyber data breaches. The major contributions of this article can be outlined as follows:
\begin{itemize}
\item[$\bullet$] This article proposes a novel User Behaviour Evaluation (UBE) concept, to scrutinize the users' intentions behind data access requests. UBE computes different crucial security parameters to conclude the security score of each data request for further analysis.
\item[$\bullet$] This research work developed a novel Federated learning-driven Malicious User Prediction (FedMUP) approach to unveil malicious users, proactively in real-time scenarios. 
\item[$\bullet$] The proposed FedMUP framework comprises mutually beneficial collaboration in federated learning scenarios to ensure enhanced data security. It furnishes individual local models for each user to obtain a global model by deploying some averaging technique to keep harmony between data sharing, implementation, privacy, and communication.
\item[$\bullet$]  Comprehensive experiments, verify that the FedMUP model can identify user classes, in a federated learning environment very efficiently and effectively with a high value of enactment parameters like accuracy, precision, recall, and F1-score in comparison to other state-of-the-art works. Further, the effectiveness of the proposed approach is achieved by operating a wide range of data sets, and feature analysis.
\end{itemize}
\begin{figure}[!htbp]%
\centering
\includegraphics[width=0.99\textwidth]{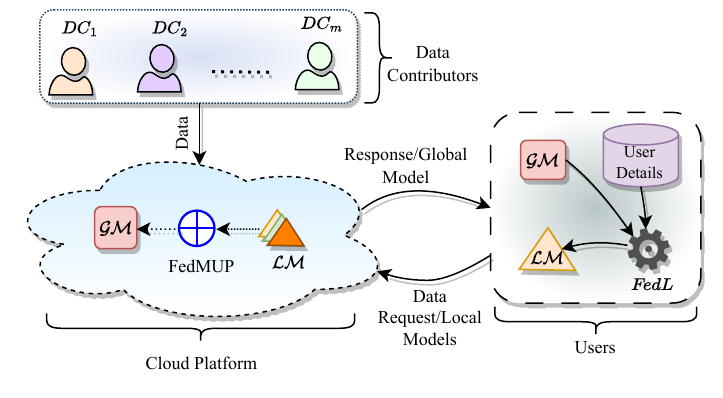}
\caption{{Operational viewpoint of the proposed FedMUP Model}}\label{opview}
\end{figure}
\par Figure \ref{opview} exhibits an operational viewpoint of the proposed FedMUP model. It is visible that the cloud platform is facilitating the storage, analysis, investigations, computations, storage, and data sharing among various stakeholders for the data supplied by the data contributors. First of all, user behavior is evaluated to confine the intention behind data access as 'malicious' or 'non-malicious'. Thereafter, the federated learning-driven malicious user prediction unit trains the local models to define the global model obtained from local model averaging. These local versions transfer weights and some other relevant details rather than actual data to ensure enhanced proactive data protection.
\par Table \ref{TableTerminology} showcases a description of the terminologies employed throughout this article.
\begin{table}[!htbp]
\footnotesize
\caption{Terminologies with descriptions}
\label{TableTerminology}
\begin{center}
\resizebox{\columnwidth}{!}{
\begin{tabular}{|ll|ll|}\hline \hline
\textit{${DC}$} & Data Contributors & $m$ & Number of Data Contributors \\
\textit{${D}$} & Data Objects & \textit{$m^\ast$} & Number of Data Objects \\
\textit{${U}$} & Users & \textit{$n$} & Number of Users \\
\textit{$\vartheta$} & Data Leakage Status & \textit{NE} & Non-entrusted Entity \\
\textit{$\zeta^{u}$} & User Historical Details & \textit{$\psi^{u}$} & User Live Details  \\ 
\textit{${DD}^{mal}$} & Malicious Data Distribution & \textit{${\varkappa}$} & Attack Factor \\
\textit{$\mu\Lambda$} & Users Attributes & \textit{${DA}^{total}$} & Total Data Access \\   
\textit{$\sigma$} & Security Parameters  & \textit{$N^\ast$} & Total Security Parameters  \\
\textit{${Thr}^{freq}$} & Threshold frequency & \textit{${\Im}$} & Data Access Type \\
\textit{$\wp$} & Security Risk Information & \textit{$\oplus$} & Averaging \\
\textit{${AD}$} & Unauthorized Data & \textit{$D_{j}$} & Requesting Data Objects \\
\textit{${DL}^{mal}$} & Data Leak frequency & \textit{$z$} & No. of Requesting Data \\ 
\textit{$\xi$} & Epoch & \textit{$T$} & Iteration/Communication round \\
\textit{$\omega$} & weight & \textit{$FedL$} & Federated Learning \\ 
\textit{$u_k$} & Participating Users & \textit{$t$} & Time \\
\textit{$\mathcal{GM}$} & Global Model & \textit{$\mathcal{LM}$} & Local Model \\
\textit{$FP$} & False Positive & \textit{$FN$} & False Negative \\
\textit{$TP$} & True Positive & \textit{$TN$} & True Negative \\
\textit{$Acc$} & Predicted Accuracy & \textit{$Prec$} & Predicted Precision \\
\textit{$Rec$} & Predicted Recall Data & \textit{$F1$} & Predicted F1-Score  \\
\hline \hline
\end{tabular}}
\end{center}
\end{table}
\par \textit{Paper organization}: The remaining organizational structure of the article is described as follows. Section \ref{secrel} comprises a concise study and encapsulation of various malicious user detection and federated learning approaches. Section \ref{secsysmod} confers the FedMUP model with basic entities involved, possible threats, challenges, goals, and basic descriptions. Also, the proposed FedMUP encompasses two separate major units a) User behavior evaluation (Subsection \ref{subsecube}), and b) Malicious user prediction (Subsection \ref{subsecmup}), by utilizing the computational capabilities of FedL-driven data security strategy. The model's operational flow, complexity, and illustration along with the implementation details are elaborated in Section \ref{secopd}. Similarly, in Section \ref{secres} all the details for the experimental setup, dataset description, performance parameters, and comparison with state-of-the-art works are highlighted. Finally, Section \ref{seccon} put forward the conclusion derived from the proposed FedMUP approach.
\section{Related Work}\label{secrel}
To clinch data security, and data privacy for secure communication ample work has been carried out by the researchers. After thorough scrutiny, a few of the models found suitable for study are listed and compared here. Correspondingly, data security can be divided into two broad categories; reactive and proactive respectively.
\subsection{Malicious Identification}\label{subsecrela}
\par A guilty agent detection model is presented by Papadimitriou et al. \cite{5487521}. This one of the pioneering model target to reveal the possible guilty agent who might be responsible for a data breach. To unveil the seem to be malicious entity model utilizes watermarking and provenance techniques to inject the fake data instance with the requested data, for easy recognition of guilty at a later stage. Though the model is highly robust it handles only explicit data requests.
\par A prompt dynamic malware detection is proposed by Rohde et al. \cite{Rohde2018}. By utilizing the ensemble recurrent neural network (RNN) model discover if the attached executable payload is malicious or not. The model performs this assigned task proactively, with a faster pace within 5 seconds hence exhibiting brilliant efficiency. Despite this model lags in performing this task across different operating systems.
\par Sharif et al. \cite{Sharif2018} devised an approach for malicious payload identification proactively. Hyper-text transfer protocol (HTTP) data of 20k users turned out over the web in 3 months duration was analyzed using machine learning classification algorithms. Extensive analysis of self-reported features, contextual features, and historical behavior features is carried out to unveil the malicious exposure. The model does not consider behavioral data for analysis purposes. 
\par Dynamic threshold-oriented data leaker identification (DT-ILIS) model is proposed by Gupta et al. \cite{gupta2019dynamic}. The access control mechanism is utilized to identify the possible data leaker by observing the pattern of data allocation among distinguished users. Model imparts a great enhancement in performance parameters like probability, success rate, detection rate, etc. The major imperfection the model suffers is that it consider the distributor as a trusted entity which is impractical in real-world scenario.
\par Lingam et al. \cite{Lingam2019} pave the way to detect social bots and influential users on social network sites. Deep Q Learning (DQL) depending on distinct social features like tweet-oriented features, user profile features, and social features is developed to detect social bots. Despite achieving high efficiency in detecting social bots and influential users, it still suffers impedance in real-time scenarios.
\par A robust approach to lookout the online information leaker (On-ILIS) is proposed by Gupta et al. \cite{guptaOnILIS}. The model focuses on a data allocation strategy to identify the leaker later. On one side, the model copes with data disclosure risk but on the other side, it snags with low data privacy issues.
\par A data security approach based on the user's behavior analysis is introduced by Rabbani et al. \cite{RABBANI2020102507}. A self-optimized learning strategy based on particle swarm optimization probabilistic neural network (PSO-PNN) is employed to map the activity pattern of various users. Though the model is capable of distinguishing the user as normal or malicious, it doesn't justify the raw data handling over high high-traffic network.
\par One of the leading probability-based malicious user identification using a machine learning approach is devised by \cite{gupta2020mlpam}. The model employs the data allocation pattern to figure out the malicious one. To add another dimension of data security and privacy, the model utilizes cipher text-based encryption along with differential privacy approaches. Proactive measure to detect guilty agent is a marvelous outcome of the model.
\par Gupta et al. \cite{gupta_Kush_2020} devised an XGBoost machine learning classification-oriented approach to impart data leakage prevention. The model introduced the ensemble XGBoost pipeline for classification purposes.
\par Afshar et al. \cite{afshar2021incorporating} came up with an attribute/behavior-oriented access control scheme (ABBAC) to detect malicious users. The model first analyzes the user behavior and accordingly, issues grants to access the data. The model has limited accuracy but performs well to impart protection from unauthorized access.
\par A highly robust and efficient privacy-preserving intrusion detection (PC-IDS) system is presented by Khan et al. \cite{Khan2021}. The model comprises two units for data preprocessing and malicious intrusion identification respectively. This model yields high performance, and detection rate but high computational complexity sets a major drawback for the model performance.
\par Raja et al. \cite {Raja2022} introduced a model to identify fake accounts existing on social platforms to ensure privacy. This model took the services of the support vector machine-neural network (SVM-NN) technique in combination with data mining approaches, to perform the task of fake account detection. The model emphasizes a few parameters such as Behavior pattern, number of shared posts, recent activities, etc. to deploy 3PS (Publically Privacy Protected System) data mining whereas SVM-NN performs classification into fake or real. The only major setback to the model is the lack of user preferences.
\par Ranjan et al. \cite {Ranjana2022} emphasized on development of novel techniques to predict malicious users on web applications. This model utilizes big data analytics, and a random forest application algorithm (RFAA) to analyze the agent behavior. The model performs malicious identification tasks with 65-70 \% accuracy thereby, giving ample scope to scale up the performance.
\par Gupta et al. \cite{9865138} offered a prominent and highly efficient malicious user prediction (MUP) model, working proactively i.e. before granting the data access to the users. To pursue malicious user detection, this model utilizes quantum machine learning (QML) employing qubits operating on Pauli gate in a multi-layer environment. This is one of the first ones to deploy QML for MUP proactively with better accuracy. The only impedance model suffers is high computational complexity. 
\par The significant observation noted here is that existing models consider only users as a non-entrusted entity though in a real scenario all the important entities like data contributors, service providers, etc. can't be trusted fully. A comparative summary of existing malicious user prediction (MUP) approaches is highlighted in Table \ref{tabrelwork1}.
\begin{table*}[!htbp]
\centering
\caption{Encapsulation of various Malicious User Prediction Approaches}\label{tabrelwork1}
\resizebox{0.96\textwidth}{!}{
\begin{tabular*}{\textheight}{@{\extracolsep\fill}llccc}
\hline \hline
Contributor & Strategy/Approach & Dataset & Implemen- & Evaluation   \\
(Timeline) & & & tation Tool & Parameters  \\
\hline \hline
Papadimitriou et & Fake objects data allocation,   & --- & Python  & Probability, \\
 al. \cite{5487521} (2011)  & Agent, Object selection &  & & Confidence detection   \\ \hline
Rhode et al.  & Malware detection based on & Virus & Python  & \textit{$Acc$}   \\
\cite{Rohde2018} (2018)  & RNN, Feedforward network & Total & 2.7 & FP, FN\\ \hline
Sharif et al. & Malicious Content Prediction, & RSeBIS & Python  & True/False   \\
\cite{Sharif2018} (2018) & Feed Forward network &  & & Positive Rate  \\ \hline
Gupta et al. & Threshold based guilt detection,  & --- & C++  & Average probabi-  \\
\cite{gupta2019dynamic} (2019) & Round robin data allocation & & & lity, Success rate  \\ \hline
Lingam et al. & Social bots, Influential users & Twitter & Python & \textit{$Acc$}  \\
\cite{Lingam2019} (2019) & detection, State-Action pairs & network & &  \\ \hline 
Gupta et al. & Online leaker detection, & UCI & Python & Average  \\
\cite{guptaOnILIS} (2021) & Data distribution & repository & & probability \\ \hline
Rabbani et al. & Malicious recognition based hybrid  & UNSW- & Python  & Precision  \\
 \cite{RABBANI2020102507} (2020) & ML, Multilayer feed-forward NN & NB15 & & F measures  \\ \hline
Gupta et al. & Data leaker identification, & UCI & Python,  & Detection, \textit{$Acc$},   \\
\cite{gupta2020mlpam} (2020) & Encryption, Probabilistic & repository & C++ & \textit{$Prec$}, \textit{$Rec$}, \textit{$F1$}  \\ \hline
Gupta et al. & Data leaker prevention, & UCI & Python  & \textit{$Acc$}  \\
\cite{gupta_Kush_2020} (2020) & XGBoost, Ensemble learning & repository & &  \\ \hline
Afshar et al. & Insider attacks detection, & UCI & Python  & RMSE  \\
\cite{afshar2021incorporating} (2021) & ABE, Analysing actual, Historical data & repository & & MRMSE \\ \hline
Khan et al. & Intrusion detection in Smart power  & UNSW- & Python  & Detection rate,  \\
\cite{Khan2021} (2021) & system (SPS), Particle swarm optimization & NB15 & & FPR  \\ \hline 
Raja et al. & Malicious profiles, Users detection, & OSN & Python  & \textit{$Acc$}   \\
\cite{Raja2022} (2021) & E$\_$SVM-NN, Data mining tools & accounts & &  \\ \hline
Ranjan et al. & Data analytics, ML-based malicious user & Magneto & MSSQL & \textit{$Acc$}  \\
\cite{Ranjana2022} (2022) & prediction, Random forest, AWS cloud  &  & & Workflow time  \\ \hline
Gupta et al. & Quantum based malicious user prediction, & CMU & Python & \textit{$Acc$}, \textit{$Prec$},  \\
\cite{9865138} (2022) & Multilayer feed forward network & CERT & & \textit{$Rec$}, \textit{$F1$} \\ \hline
\textbf{This Work} & Malicious user prediction, & CMU CERT & Python  & \textit{$Acc$}, \textit{$Prec$},  \\
\textbf{(FedMUP)}
& using FedL through ANN,CNN,DNN & r4.2 & 3.6.8 & \textit{$Rec$}, \textit{$F1$}, loss \\ 
\hline \hline
\multicolumn{5}{p{19cm}}{$Acc$: Accuracy; $FP$: False Positive; \textit{$FN$}: False Negative; \textit{$Prec$}: Precision; \textit{$Rec$}: Recall; \textit{$F1$}: F1-Score; \textit{$FPR$}: False Positive Rate;  \textit{$RMSE$}: Root Mean Square Error;}
\end{tabular*}
}

\end{table*}
\subsection{Federated Learning Approaches}\label{subsecrelb}
In recent years federated learning approach has emerged as a prominent and effective way to impart data security and data privacy. It overcomes the leading issue with existing model training approaches i.e., the privacy breach due to data gathering at a centralized location for model training. Federated learning brings the training model to a local location rather than asking for private data at Central premises.
\par Asad et al. \cite{ASAD2021107235} has proposed a novel comprehensive Federated Learning-based Communication Efficient and Enhanced Privacy (CEEP-FL) approach to mitigate the issue of a private data breach due to data sharing over centralized location. This model deploys a threshold value-based filtering strategy to extract only important local gradients, and a Non-Interactive Zero-Knowledge Proofs based Homomorphic-Cryptosystem (NIZKP-HC) for encrypting local updates to share over cloud server, followed by Distributed Selective Stochastic Gradient Descent (DSSGD) optimization to enhance learning accuracy of global version. The model successfully imparts low computational cost and high data protection with robustness but with the scope of growing further.
\par Another, federated learning-based recommender system is proposed by Du et al. \cite{DU2021107700} by considering the recommender as untrusted. This model defines a user-level distributed matrix factorization by gathering gradients, not raw data from users. The model yields high privacy with lower computations by using homomorphic encryption. A most important consideration to work here is that the model considers all users at par which seems practically not true.
\par An optimized poisoning attack with persistence and stealth features model is presented in \cite{zhou2021deep}, which protects the data by adding malicious neurons into the global model. The authors enhanced the durability, efficacy, and robustness of poisoning attacks by injecting malicious neurons in the redundant network space using the regularization term in the objective function. The proposed model achieved the highest attack success rate and accuracy. A limitation of this model is that it has limited data sharing.
\par A federated learning framework that analyses the common aggregation strategy was proposed by Mansour et al. \cite{mansour10069463}. Additionally, the authors designed a group of hybrid techniques that combine FedAvg and aggregation algorithms and can change the architecture by varying client contributions based on the magnitude of their losses. The proposed framework does not provide data protection but increases stability and efficiency.
\par Xu et al. \cite{XU2022109488} developed a federated learning-driven scenario to cope with the problem of data shortage due to security concerns. This scheme proposed FedFSL for few-shot learning, WFedL to balance the training randomness and data distributions for improved performance parameters and CSFedL to eradicate malicious participation during training and global model construction. Though the model outperforms concerning performance parameters, it still lags in terms of real data application.
\par A personalized federated learning model via mutually beneficial collaboration (FedMBC) was proposed by Goang et al. \cite{GONG2023}. It gives each client access to an aggregated model by fostering cooperation among customers with similar needs. To aggregate a model suitable for its local data distribution, the authors developed a dynamic aggregation method based on the similarity of clients on the server in each communication round. The proposed model balances communication, performance, and privacy with some limits on data sharing.
\par An effective Decentralized Federated Learning Historical Gradient (DFedHG) approach was devised to identify and separate trustworthy, malevolent, and normal users \cite{CHEN2023105888}. The developed algorithm prevents data breaches by classifying the users. Although the proposed method preserved data privacy, it has less efficiency and classification accuracy. A comparative summary of existing federated learning-based malicious user identification approaches is highlighted in Table \ref{tabrelwork2}.
\begin{table*}[!htbp]
\caption{Encapsulation of various Federated Learning Approaches}\label{tabrelwork2}
\resizebox{\columnwidth}{!}{
\begin{tabular*}{\textheight}{@{\extracolsep\fill}llccc}
\hline \hline
Contributor & Strategy/Approach & Dataset & Implemen- & Evaluation  \\
(Timeline) & & & tation Tool & Parameter \\
\hline \hline
Asad et al.  & An efficient and protected & CIFAR-10 & Python  & Computational  \\
\cite{ASAD2021107235}(2021) &  
 communication, \textit{HE}, \textit{SGD} & & Tensor flow & time, Accuracy \\\hline 
Du et al. & User-level distributed matrix & Filmtrust & Python  & Accuracy \\
\cite{DU2021107700} (2021) &  factorization, \textit{HE}, FedML & Movielens & & \\ \hline
Zhou et al. & An optimized poisoning  & MNIST,  & Python & Accuracy  \\
 \cite{zhou2021deep}  (2021) & attack, CNN, FedML &  CIFAR-10 &  & \\
\hline
Mansour et al. & Federated learning & MNIST,  & Python & Accuracy \\
\cite{mansour10069463}  (2022) & aggregation, CNN & Fashion\_MNIST &  &  \\
\hline
Xu et al. & A joint training data & CIFAR-10 & Python & Training run \\
\cite{XU2022109488} (2022) & protection, FedSFL, WFedL & CIFAR-100 & &  time, Accuracy  \\ \hline
Goang et al.  & Personalized mutually beneficial colla- & MNIST,  & Python & Computational \\
\cite{GONG2023} (2023) & boration, \textit{DA}, Local update, FedML & CIFAR-10, 100 &  & cost, Accuracy \\  \hline
Chen et al. & Historical gradient based Non-  & MNIST,  & Python & Accuracy, \\
\cite{CHEN2023105888} (2023) & trust detection, FedML & FMNIST &  & Loss \\
\hline \hline
\end{tabular*}}
\end{table*}
\par One of the common observations recorded from the above-stated studies is that the model needs data for training purposes but due to frequent data leakages sensitive data become prone to mishandling. As a result, the data owner restricts data sharing, data access, and data availability. Data is the key to expansion. The need of the hour is to make data sharing available with utmost access and protection but not at the cost of data availability. Federated learning does not necessitate data sharing rather performs model training locally and shares the model, not the raw data. Thereby, it appears as a prominent way to ensure learning but not at the cost of data availability and data security.
\section{{Proposed FedMUP Model}}\label{secsysmod}
This section portrays all the entities participating in the model with their allotted duties, describes the crisis, and recaps the design purposes. The system model incorporates three significant entities: \textit{Data Contributors (DC)}, \textit{Cloud Platform (CP)}, and the \textit{Users (U)} which are explained as follows:
\begin{enumerate}
    \item \textit{Users (U)}: An entity that raises a request to access the data and receives the same from the \textit{CP}, with the intent of some utility. The system model believes \textit{U} as a non-entrusted entity. 'Malicious', 'non-malicious', and, 'unknown' are three distinct classes into which \textit{U} can be categorized.
    \item \textit{Data Contributors (DC)}: An entity, responsible for the contribution of data to the \textit{CP} for further services. \textit{DC} gathers the \textit{data objects} \{${D_1}$, ${D_2}$, ${D_3}$, ..., ${D_{m^\ast}}$\} and share these with \textit{CP} for analysis, computation or storage purposes. The system model assumes it is a non-entrusted entity because of the high possibility of data leakage itself.
    \item \textit{Cloud Platform (CP)}: An entity that acts as an interface between \textit{DC} and \textit{U}. It collects the data objects shared by the data contributor, stores these to facilitate computational, analysis or data sharing, and finally supplies them to the requesting users. Before furnishing users' requests, it employs the UBE (Section \ref{subsecube}) and Malicious user prediction (Section \ref{subsecmup}) to outline the possible malicious user, proactively. The system model considers it as a non-entrusted entity as it might be curious to remember the details.
\end{enumerate}
\par The model presumes that all the entities such as \textit{DC}, \textit{CP}, or \textit{U} are authorized-but-curious adversaries. However, all the entities seek the model instruction, still a high chance of data security compromise exists as any entity can be directly or indirectly responsible for the leakage of the data. Particularly, a malicious user may get data accessed and hence the confidential information gets revealed. Moreover, users need to share private cum sensitive data during central model training purposes and model updates. In this scenario, malicious users, cloud platforms, or even third parties may infer the sensitive information security approach hence making ground for the need for an efficient data security approach. 
\par
\begin{figure}[!htbp]%
\centering
\includegraphics[width=0.99\textwidth]{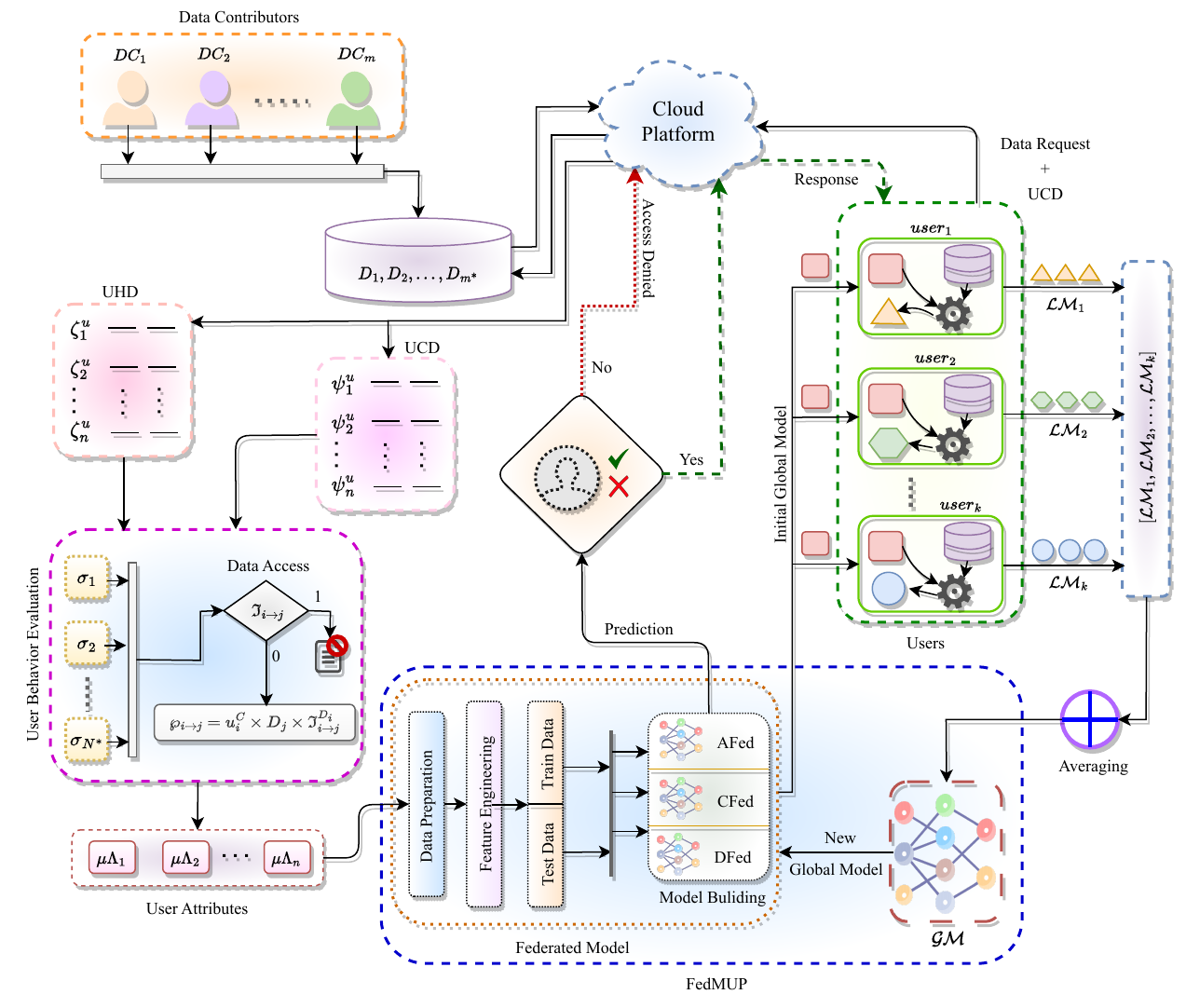}
\caption{Architecture of proposed FedMUP Model}\label{architecture}
\end{figure}

 Various \textit{DC}, comprising data objects (\textit{D}) are required to share data with \textit{CP} to fulfill the \textit{U} data access requests, required for the growth of an organization. Prominent challenges are data security and privacy from a malicious user or data compromise during central model training or modal updates without causing performance degradation. The primary goal is to impart a proactive, malicious user prediction approach. Another significant goal is to conserve data confidentiality by way of the model-to-data rather than a data-to-model approach for enhanced data security. To formulate an approach to resolve the issue of performance degradation.
The architecture of the proposed FedMUP Model has been illustrated in Figure \ref{architecture}. A detailed view of notable entities and crucial blocks partaking, along with interaction and elementary flow among these entities and blocks, is depicted.
\par Study assumes that the cloud platform \textit{CP} receives data access request from participating $k$ users: \{$u_1$, $u_2$, $u_3$, ..., $u_k$\} at a time, out of existing $n$ \textit{users}: \{$u_1$, $u_2$, $u_3$, ..., $u_k$\} $\in {U}$. Here, $m$ \textit{data contributors}: \{${DC}_1$, ${DC}_2$, ${DC}_3$, ..., ${DC}_m$\} $\in {DC}$ are sharing a total of $m^\ast$ \textit{data objects}: \{${D}_1$, ${D}_2$, ${D}_3$, ..., ${D}_{m^\ast}$\} over the \textit{CP} for the purpose of storage, computation and most importantly data sharing among $n$ users: \{$u_1$, $u_2$, $u_3$, ..., $u_n$\} $\in {U}$ to accomplish their request for data access. \textit{Users' current details}: ${\psi}^u$ = \{${\psi}_1^u$, ${\psi}_2^u$, ${\psi}_3^u$, ..., ${\psi}_n^u$\} along with \textit{User's historical details}: ${\zeta}^u$ = \{${\zeta}_1^u$, ${\zeta}_2^u$, ${\zeta}_3^u$, ..., ${\zeta}_n^u$\} associated with data access requests are supplied into (UBE) unit to estimate the user behavior.
\par UBE evaluates multiple ($N^\ast$) \textit{security risk parameters}: $\sigma$ = \{${\sigma}_1$ $\cup$ ${\sigma}_2$ $\cup$ ${\sigma}_3$ $\cup$ ... $\cup$ ${\sigma}_{N^\ast}$\} associated with all $n$ users request to fetch the \textit{security risk information}: $\wp$ = \{$\wp_1$, $\wp_2$, $\wp_3$, ..., $\wp_n$\} (additional details in Section \ref{subsecube}). Finally, it computes the values of \textit{users attributes}: $\mu\Lambda$ = \{$\mu\Lambda_{1}$, $\mu\Lambda_{2}$, $\mu\Lambda_{3}$, $\dots$, $\mu\Lambda_{n}$\}. The dataset \cite{dataset} utilized for the purpose of user behavior evaluation is comprised of various historical and current user details. Based on these user details, user behavior is evaluated. The users' current details ($\psi^u$), historical details ($\zeta^u$), and computed user attributes ($\mu\Lambda$) are stored in a knowledge database for model re-training purposes. The FedMUP model employs these for further analysis to accomplish the purpose of proactive malicious user detection.
\par This unit takes benefit of the high computational power of deep learning computations and machine learning concepts. The proposed FedMUP model obtains and examine \textit{users' attributes}: $\mu\Lambda$ = \{$\mu\Lambda_{1}$, $\mu\Lambda_{2}$, $\mu\Lambda_{3}$, $\dots$, $\mu\Lambda_{n}$\} alongwith multiple \textit{security risk values}: $\wp$ = \{$\wp_1$, $\wp_2$, $\wp_3$, ..., $\wp_n$\} in respect of various users, received from UBE. As portrayed in Figure \ref{architecture}, federated learning-based deep machine learning approaches like Artificial Neural Network-driven federated learning (AFed), Convolutional Neural Network-driven federated learning (CFed), and Deep Neural Network-driven federated learning (DFed) are utilized to investigate and predict the possibility of malicious activity before the allocation of requested data in a real-time cloud communication environment. As described in Section \ref{subsecmup} basic initial global model is generated on behalf of these calculated values which is further shared with the different participating \textit{users}: \{$u_1$, $u_2$, $u_3$, ..., $u_k$\} $\in {U}$ for training and learning over their local datasets without need of data sensitive transfer from the local device to a central server and hence releases local models obtained from each of the participating or eligible users.
\par These \textit {local models}: $\mathcal{LM}$ = \{$\mathcal{LM}_1$, $\mathcal{LM}_2$, $\mathcal{LM}_3$, ..., $\mathcal{LM}_k$\} are than combined using averaging techniques to obtain a new enhanced version of the global model; a more updated, informed, and efficient one. Hence, the Federated Learning (FedL) implementation cycle comprises model selection, model sharing, local model training, and aggregation of the local model to formulate a new updated global model. This process of learning, sharing, and updating is repeated for the next iteration and so on till the desired level of performance is achieved to detect the possible malicious user with high accuracy. For outcomes determined as non-malicious by this unit; data is distributed among requesting users keeping high security, otherwise, they are denied. Thus FedMUP model demonstrates a potent strategy for data security problems. It considers all the participating entities as non-entrusted thereby providing a composite mechanism to protect data proactively in a sharing environment. The detailed description of User Behavior Evaluation and Malicious User Prediction is performed in Subsection \ref{subsecube} and Subsection \ref{subsecmup}, respectively.
\subsection{{User Behaviour Evaluation}}\label{subsecube}
Suppose that the $CP$ receives the data access request from various $U$ for data objects $\{D_1, D_2, D_3,..., D_{m^\ast}\}$. The FedMUP model ensures the users' behavior evaluation proactively for being 'malicious' or 'non-malicious', based on the current ($\psi^u$) and historical details ($\zeta^u$) of the requesting user. Table \ref{dataset} showcases a brief view of these current and historical details of the requesting user. Considerable security parameters comprising various important values such as \textit{users' historical details} ($\zeta^{u}$); \textit{users' authorized set of data units} (${AD}$); total number of \textit{ malicious data distributions or data leakages} (${DD}_i^{mal}$) performed by user $u_i$ over a specified time-interval \{$t_\alpha$, $t_\beta$\}; and the \textit{data breaches frequency} (${DL}^{mal}$) are assessed to evaluate the user intentions. The mathematical formulation and computation for security parameters ($\sigma$) are described as follows:
\begin{itemize}
\item \textit{Users' historical details} ($\zeta^{u}$): To find out if requesting user $u_i$ is already a `known' or `unknown', record of the data access in the past by user i.e., $\zeta^{u}$ is computed by employing Eq. (\ref{sp1}). Here, ${\sigma}_{\zeta^{u}_i}$ denotes the security parameter ($\sigma$) related with users' historical details $\zeta^{u}_i$. For a known user, some historical details exist in the knowledge base which help to check the susceptibility of the requesting user. 
\begin{gather} \label{sp1}
{\sigma}_{\zeta^{u}_i} = 
 \begin{cases}
  \textit{Known} \; (0),  & \text{If}  (|\zeta^{u}_i| > 0)\\
 \textit{Unknown} \; (1), & Otherwise\\
 \end{cases}
 \end{gather}

\item \textit{Authorized data for user} (${AD}$):  Eq. (\ref{sp2}) compute that is requesting user $u_{i}$ is authorized to access the data (${AD}_i$). Here, $q_{1}$, $q_{2}$, $\dots$, $q_{n^\ast}$ defines the number of dataset of categories: ${C}_{1}$, ${C}_{2}$, $\dots$, ${C}_{n^\ast}$, respectively. It is considered that, $i^{th}$ user is allowed to access these data. Correspondingly, the security parameter (${\sigma}_{{AD}_i}$) associated with data access eligibility is presented in Eq. (\ref{sp3}).
\begin{gather}\label{sp2}
{AD}_i = ({C}_{1} \times \sum_{k=1}^{q_{1}}D_{k}) \cup ({C}_{2} \times \sum_{k=1}^{q_{2}}D_{k}) \cup \dots \cup ({C}_{n^\ast} \times \sum_{k=1}^{q_{n^\ast}}D_{k})
 \\
  \label{sp3}
{\sigma}_{{AD}_i} = 
 \begin{cases}
 \textit{Authorized} \; (0),  & \text{If} \; (u_{i} \times ({C}_{i} \times D_{i}) \subseteq {AD}_{i})\\
  \textit{Unauthorized} \;  (1), & Otherwise\\
 \end{cases}
 \end{gather}

\item \textit{Malicious data distributions} (${DD}_i^{mal}$): Assume that the data \{$D_{1}$, $D_{2}$, $D_{3}$, $\dots$, $D_{z}$\} is requested by the user $u_{i}$ during time-interval \{$t_{\alpha-1}$, $t_{\beta-1}$\}. Eq. (\ref{sp4}) evaluates the total number of malicious data distributions (${DD}_i^{mal}$). Here, the leakage status ($\vartheta_{i}$) of the data is assessed as $\vartheta_{i}$ = \{0, 1\} which can be either true (1) or false (0). 
\begin{gather} \label{sp4}
    {DD}_i^{mal} = \sum_{i=1}^{z}(D_{i} \times \vartheta_{i} \times t) \quad  \forall_t \in \{t_{\alpha-1}, t_{\beta-1}\}
\end{gather}  
Eq. (\ref{sp5}) evaluate the attack factor ($\varkappa$) for the user $u_{i}$. Here, ${DA}^{total}$ represents the \textit{total data accesses} during time period \{$t_{\alpha-1}$, $t_{\beta-1}$\}. Further, whether the data access request of any user $u_i$ should be fulfilled or rejected is determined by Eq. (\ref{sp6}), by utilizing the malicious data distributions (${DD}_i^{mal}$) performed in past. Here, $Thr^{attack}$ has to have a predefined value of 0.5 considered based on the experimentation work. 
\begin{gather}
 \label{sp5}
\int_{t_{\alpha-1}}^{t_{\beta-1}} {\varkappa}_{i}\; dt = \int_{t_{\alpha-1}}^{t_{\beta-1}} \frac{{DD}_i^{mal}}{{DA}_i^{total}} dt
\\ \label{sp6}
{\sigma}_{{\varkappa}_{i}} = 
 \begin{cases}
  \textit{Access allowed} \; (0),  & If \; (Thr^{attack} > {\varkappa}_{i})\\
  \textit{Access denied} \; (1), & Otherwise\\
 \end{cases}
 \end{gather}
\item \textit{Frequency of data breaches} (${DL}^{mal}$): Eq. (\ref{sp7}) evaluates the probability for malicious data leakage (${DL}^{mal}$). Here, $\sum\limits_{k=1}^{H} D_{z_k} \not\in {AD}_{i}$ and $t_{ijk}$ specifies the number of times the $i^{th}$ user $u_{i}$ has tried to access non authorized data ($D_{z_k}$) over $j^{th}$ time-period. Also, $H$ and $T$ describe the total number of unauthorized data requests submitted by the user $u_{i}$ in time duration $T$ wherein, $T$ $\in$ \{$t_{(\alpha-1)}$, $t_{(\beta-1)}$\}. Further, whether to grant or reject data access to the user $u_{i}$, is computed using Eq. (\ref{sp8}). Accordingly, it allows data access to $D_{j}$ only in case ${\sigma}_{{DL}^{mal}}$ = 0 else access to data is denied. Here, $Thr^{freq}$ represents the threshold frequency of unauthorized data access attempts in the time period \{$t_{(\alpha-1)}$, $t_{(\beta-1)}$\}. It is required to be as minimum as possible. Therefore, a predefined value of 0.3 is considered more suitable. 
\begin{gather}
\label{sp7}
    {DL}^{mal}_{i} = |\sum_{k=1}^{H}\sum_{j=1}^{T} D_{Z_k} \times t_{ijk} \times u_{i}|\\
    \label{sp8}
{\sigma}_{{DL}^{mal}} = 
 \begin{cases}
\textit{Allowed} \; (0),  & If \; (Thr^{freq} > {DL}^{mal}_{i})\\
  \textit{Denied} \; (1), & Otherwise\\
 \end{cases}
 \end{gather}

\end{itemize}

\par Ultimately, all computed  ${N^\ast}$ security parameters ($\sigma$) are aggregated as shown in Eq. (\ref{final1}). Moreover, Eq. (\ref{final2}) assess the intention behind data access where $\Im_{i \rightarrow j}^{D_{i}}$ decides data access type as non-malicious ($\Im_{i \rightarrow j}^{D_{i}}$ = 0) or malicious ($\Im_{i \rightarrow j}^{D_{i}}$ = 1) over the time-period \{$t_{\alpha}$, $t_{\beta}$\} by utilizing Eq. (\ref{final3}).
\begin{gather}
\label{final1}
{\sigma} ={\sigma}_{\zeta^{u}_i} +{\sigma}_{{AD}_i}+{\sigma}_{{\varkappa}_{i}}  +{\sigma}_{{DL}^{mal}}+ \dots + {\sigma}_{N^\ast}\\
\label{final2}
    \Im_{i \rightarrow j}^{D_{i}}  = 
      \begin{cases}
           \textit{Non-malicious} \; (0),  & If \; ({\sigma} < 1)\\
            \textit{Malicious} \;(1), & Otherwise\\
        \end{cases}
        \\ 
        \label{final3}
       \int_{t_{\alpha}}^{t_{\beta}}\wp_{i \rightarrow j} dt = \int_{t_{\alpha}}^{t_{\beta}} (u_{i}^{{C}} \times D_{j} \times \Im_{i \rightarrow j}^{D_{i}})\; dt  
        \end{gather}     \\   

In this way, the proposed model determines the intention of the user behind the raised data request, proactively.
\subsection{{Malicious User Prediction}}\label{subsecmup}
The proposed model under consideration predicts malicious users proactively, to provide up-scaled data protection. A few essential actions described further are performed before model training. First of all, data pre-processing is accomplished for further processing, followed by participating user selection for the task of model training, and ultimately performing local and global model training in a Federated Learning (FedL) environment.
\begin{itemize}
\item \textbf{Data Preprocessing:} UBE provides the various users' details \{$\mu\Lambda_{1}$, $\mu\Lambda_{2}$, $\mu\Lambda_{3}$, $\dots$, $\mu\Lambda_{n}$\} $\in \mu\Lambda$ and numerous security risk values \{$\wp_1$, $\wp_2$, $\wp_3$, ..., $\wp_n$\} $\in$ $\wp$ to $CP$ for data preparation. Subsequently, preprocessing with the intent of data cleaning to overcome missing data (if any), and to avoid any kind of data mishandling which further might hinder the models' performance is performed. The obtained data values are normalized or re-scaled in the range [0,1] using Eq. (\ref{sp12}).
\begin{gather} \label{sp12}
{\hat{\mu\Lambda}_j} = \frac{{\mu\Lambda}_j - {\mu\Lambda}_{min}}{{\mu\Lambda}_{max} - {\mu\Lambda}_{min}}
\end{gather}
\par wherein ${\mu\Lambda}_{max}$ and ${\mu\Lambda}_{min}$ are the maximum and minimum values of the input data set. Further, the preprocessed data \{${\hat{\mu\Lambda}}_1$, ${\hat{\mu\Lambda}}_2$, ${\hat{\mu\Lambda}}_3$, ..., ${\hat{\mu\Lambda}}_n$\} is bifurcated in 80:20 ratio for model training and testing, respectively.
\item \textbf{User Participation:} All $U$ do not necessarily raise data access requests simultaneously. Users' participation depends on their availability, gradient data sensitivity, and relevance of data. The presented model assumes that $k$ users out of $n$ users are participating at a moment.
\end{itemize}
\begin{figure}[!htbp]%
\centering
\includegraphics[height=8.5cm, width=0.9\textwidth]{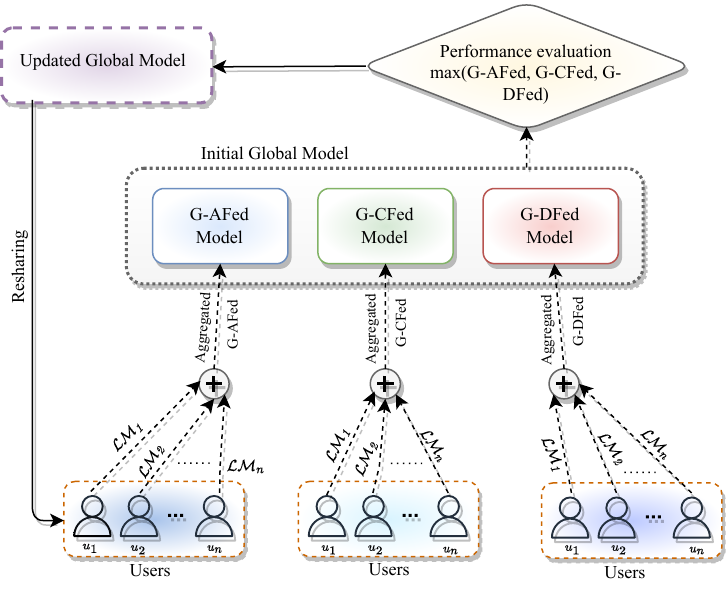}
\caption{Model Selection Approach}\label{modelselec}
\end{figure}
\par The proposed FedMUP model accomplishes malicious user prediction tasks through the continuous iterative learning process and its enactments are composed of a few repetitive actions for the availability of the latest updated global model ($\mathcal{GM}$) up-gradation across all users. Figure \ref{modelselec} visualizes the FedMUP selection process with its workout technique.
\begin{itemize}
   \item \textbf{Model selection:} Firstly, the global pre-trained global FedL model ($\mathcal{GM}$) using various deep learning approaches like AFed, CFed, and DFed along with its basic initial parameters is shared with all the participating users ($u_k$) in the FedL environment. The best-performing version is picked for sharing.
    \item \textbf{Local model training:} Afterwards, all the participating users ($u_k$) train the local version of the model on individual sensitive data using the shared initial global model and its initial parameters, named local FedL models ($\mathcal{LM}$).
    \item \textbf{Aggregation of local models:} The obtained local FedL models ($\mathcal{LM}$) are shared with the cloud platform to accomplish aggregation in order to obtain the latest updated global FedL model ($\mathcal{GM}$). This updated global FedL model is again shared among users ($u_k$) for further learning in repetitive iterations.
\end{itemize}
   \par In the FedMUP model, every user trains their local model ($\mathcal{LM}$) on their private dataset and shares the learned parameters rather than data itself with the $CP$ providing the services. The global model ($\mathcal{GM}$) is obtained by aggregation ($\oplus$) of these local model parameters as represented in Eq. (\ref{sp13}). Initial global model parameters can be defined as $n$ is the total number of users, $k$ is the number of users participating, $T$ is the number of communications rounds i.e. iterations, $\xi$ is the number of epochs for the local model training for each communication round, $\omega^k_{t+\xi}$ represents the local model parameters of $k$ participants users, and $\omega_{t+\xi}$ represents the global model parameters. Models are described in terms of weights assigned after the completion of every epoch $t \in \{0, ..., (T\xi-1)$\} for the local model as $\omega^i_{t +\xi}$ and final weights $\omega_{t +\xi}$ for the global model after each communication round/iteration $t \in \{0, \xi, 2\xi, ..., (T-1)\xi$\}. Eventually, the service provider aggregates all these parameters to compute the latest model. This process is repeated continuously till the performance keeps upgrading.
\begin{gather} \label{sp13}
{\omega_{t+\xi}} \leftarrow \sum_{i=1}^{k}\frac{{n}_k}{n}\omega^k_{t+\xi}
\end{gather}
\par The outcomes of the proposed FedMUP model are defined on the basis of the following metrics: \textit{True Positive (TP):} Total number of users detected as malicious, which are actually malicious. \textit{True Negative (TN):} Total number of users detected as non-malicious users, which are actually non-malicious users. \textit{False Positive (FP):} Total number of users detected as malicious users, which are actually non-malicious users. \textit{False Negative (FN):} Total number of users detected as non-malicious, which are actually malicious users. Further, various performance metrics Acc, Prec, Rec, and F1 are computed in Eqs. (\ref{sp14})-(\ref{sp17}) to compare FedMUP model with state-of-the-art works.
\begin{gather}
\label{sp14}
{Acc} =\frac{TN+TP}{TN+FP+FN+TP} \\
\label{sp15}
{Prec} =\frac{TP}{TP+FP} \\
\label{sp16}
{Rec} =\frac{TP}{TP+FN} \\
\label{sp17}
{F1} =2\times\frac{Prec\times Rec}{Prec+Rec}
\end{gather}
\par Thus, the proposed FedMUP model has immense capability to train models collaboratively. Moreover, the FedMUP model builds a global model, by utilizing local model parameters, learned from training on their real data. In this way, the FedMUP model ensures the privacy, data protection, and prediction of malicious users.
\section{Algorithm and Complexity Computation}\label{secopd}
\begin{algorithm} [!htbp]
\caption{\textbf{FedMUP: Operational Pathway}}\label{algo1}
\begin{algorithmic}[1]
\STATE \textbf{Initialize:} \textit{Users} list ($U$) with correlated attributes and requested \textit{data objects:} ${D}_1$, ${D}_2$, ${D}_3$, ..., ${D}_{m^\ast}$ \; \\
\STATE \textbf{Input:} \textit{Number of users, number of selected users, number of communication rounds} and, \textit{number of epochs} ($n$, $k$, $T$, $\xi$) \; \\
\STATE \textbf{Output:} Global model ($\omega_{T+\xi}$) obtained from local models aggregation of each participating user \{$\omega^1_{t +\xi}, ..., \omega^k_{t+\xi}$\} \; \\
\STATE Periodical, training and re-training of FedMUP, with users' current ($\psi^u$) and historical information ($\zeta^u$) \; \\
\FOR{each time-interval \{$t_{\alpha}$, $t_{\beta}$\}}
    \FOR{each user ($u_i: i \in$ \{1, 2, 3, ..., n\})}
        \STATE Receive and analyse user requests to obtain security risk information \; \\
          $\sigma$=\{${\sigma}_1 \cup {\sigma}_2 \cup {\sigma}_3 \cup ...\cup {\sigma}_{N^\ast}$\}\;
        \STATE Examine the possible intention behind data access request by Eq. (\ref{final3})\; \\
        \STATE \textbf{Initialize:} $\omega_0$
        \FOR{each communication round $t \in \{0, \xi, 2\xi, ..., (T-1)\xi$\}}
            \STATE $I_t = $ set of participating $k$ users \; \\
            \FOR{each user $i \in I_t$}
                \STATE Input: security parameters and user attributes ($\wp$, ${\mu\Lambda}$)\; \\
                \STATE Local model training using without actual local data sharing \; \\
                \STATE $\omega^i_{t +\xi} = $ local model updates ($i$, $\omega_t$) \; \\ 
            \ENDFOR
            \STATE $\omega_{t +\xi}$ $\leftarrow$ $\oplus$ ($\omega^1_{t +\xi}$, $\omega^2_{t +\xi}$, $\omega^3_{t +\xi}$, ..., $\omega^k_{t+\xi}$) i.e. global update in Eq. (\ref{sp13}) \; 
        \ENDFOR
        \STATE Return: global update ($\omega_{t +\xi}$) 
        \STATE $\wp^{Int} \leftarrow$ {FedMUP(${\mu\Lambda}$, $\wp$, $\omega_{t +\xi}$)} \; \\
        \IF{$\wp^{Int} $ $>$ $0$}
            \STATE {`Malicious': user $u_i$}
            \STATE {Data access not granted \;}\;
        \ELSE
            \STATE {`Non-malicious': user $u_i$}
            \STATE {Data access granted\;}\;
            \STATE {Requested data ${D}_i$ distributed to user}
            \ENDIF
    \ENDFOR
\ENDFOR
\end{algorithmic}
\end{algorithm}
\par Algorithm \ref{algo1} signifies the fundamental execution of the proposed FedMUP model. Different users with their correlated attributes, and corresponding data requests, are initialized. Thereafter, various parameters of the proposed model such as the number of users participating $k$, epochs $\xi$, and iterations $T$, etc. are defined. Model training and re-training keep on executing, persistently for $\textit{t}$ intervals to determine whether to grant data access or deny the request, through steps (4-30). More significantly, local model training is performed for different iterations $T$ and epochs $\xi$ to obtain an aggregated global model without driving data sharing, consequently resulting in high data security. This algorithm successfully conveys the point that the proposed FedMUP model is working proactively to fulfill defined data security, and data availability without performance degradation for highly protected cloud platform communication.
\subsection{Complexity Computation}\label{subsecopda}
Algorithm \ref{algo1} furnishes the step-by-step layout of the operational flow of the model under consideration. Miscellaneous basic functions such as initialization of user list, data requests, weights, epochs, iterations, etc. in steps (1-3) consume ${O}(1)$ complexity. Step (4-30) carries out periodical training and retraining on the model where complexity depends on the size of network $L$, epochs $\xi$, and training samples $n$ i.e. ${O}(n\times{\xi}\times$L$)$. Steps (5-30) iterate over $t$ time-interval, for all users. In steps (7-8) (${N}^{\ast}$) security parameters ($\sigma$) for each user are analyzed to find out intentions behind data access, show complexity ${O}(N^{\ast})$. Steps (9-19) compute the local and global model updates. Steps (20-27) examine and grant or deny data access accordingly causing ${O}(1)$ complexity. This model comprises of all-over complexity to be ${O}(ntL\xi{N}^{\ast})$.
\subsection{Illustration}\label{subsecopdc}
\begin{figure}[!htbp]%
\centering
\includegraphics[width=1.0\textwidth]{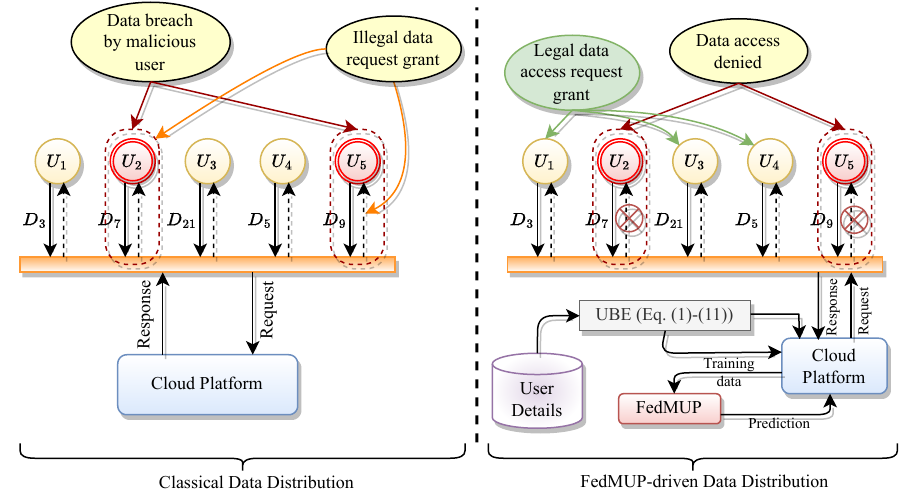}
\caption{Illustration for Classical vs FedMUP Model}\label{illus}
\end{figure}

It is assumed that the presented model is possessing five \textit{users} \{$u_1$, $u_2$, $u_{3}$, $u_{4}$, $u_{5}$\} which have raised request for \textit{data objects} \{${D}_3$, ${D}_7$, ${D}_{21}$, ${D}_{5}$, ${D}_9$\}, respectively. It is assumed that the users, $u_2$ and $u_5$ are malicious and are looking for confidential data (${D}_{7}$, ${D}_{9}$) with the intent to breach this data. This scenario is well depicted in Figure \ref{illus}.
\par Users demand to access the data and demanded data is allocated between demanding users employing existing data sharing strategies. In the classical data security approaches, demanded data is distributed without evaluating users' intentions regarding data utility. This might lead to the grant of non-authorized data access for the requested dataset ${D}_{7}$, ${D}_{9}$ and at a later stage it emerges in the form of a data breach, and consequently the hunt trigger to identify the hostile user. Rather, the presented FedMUP model serves proactively, to scrutinize the users and their several attributes furnished with demand raised for data access. The proposed model classifies the intention for access to data for every user and hence classifies users: \{$u_1$, $u_{3}$, $u_{4}$\} as non-malicious and the users: \{$u_2$, $u_{5}$\} as malicious. Therefore, in this way, FedMUP ensures enhanced proactive data security and secure communication through early identification of potential data leakages and thereby rejecting or accepting the data access demand, accordingly.
\subsection{Implementation}\label{subsecopdb}
The design and operational flow of the proposed FedMUP model is depicted in Figure \ref{implementation}. Precisely, a collaborative work of various modules and procedures eventually forms this model. A concise description of the same is specified here:
\begin{figure}[!htbp]%
\centering
\includegraphics[height=6cm, width=1.005\textwidth]{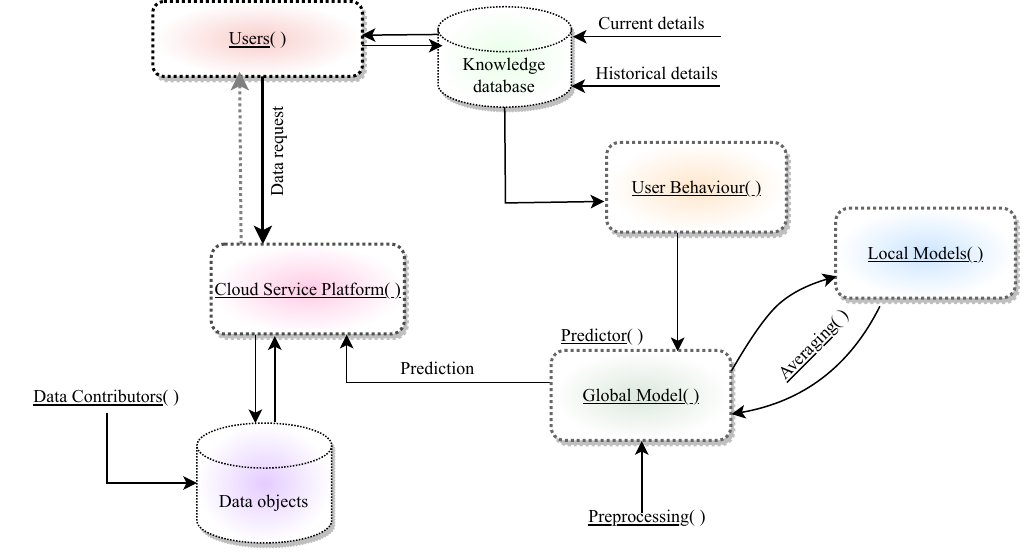}
\caption{Design and Operational flow}\label{implementation}
\end{figure}
\begin{itemize}
    \item \textit{Users()}: Users come to the Cloud Service Platform() regularly, for data access requests required for some utility purpose. It also furnishes the current and historical details along with the request for further analysis.
    \item \textit{Data Contributors()}: The Data Contributors() supply the data to the Cloud Service Platform() for ease of growth. Data accumulation is an ongoing and continuous process.
    \item \textit{Cloud Service Platform()}: It acts as a facilitator for data storage, sharing, computation, and analysis services. Data is collected from Data Contributors(), supplied to Users() as per their demand along with analysis service for ensuring data security. 
    \item \textit{Preprocessing()}: This module performs the basic data cleaning tasks such as normalization, data encoding, text-to-numeric data conversion, etc. to adapt data suitable to machine learning algorithms format.
    \item \textit{User Behaviour()}: To ensure proactive detection of malicious users, UBE (\ref{subsecube}) performs the user behavior analysis based on associated current and historical details of users.
    \item \textit{Predictor()}: The core task of the model is the prediction of the malicious user using federated learning concepts for enhanced data security and performance up-gradation by local and global model training. Predictor() receives user attributes ($UA$) from UBE to conduct further analysis.
    \item \textit{Local Models()}: These are local versions of models trained at the user site using model-to-data concepts. 
    \item \textit{Global Model()}: The model is obtained from averaging ($\oplus$) of the Local Models() of participating users. This model is again shared with local users for training over their local data without seeking data at a central location.
    \item \textit{Averaging()}: The technique to drive the final Global Model() from various Local Models() considering performance up-gradation.
    \item \textit{Knowledge database}: A data repository comprising user details for current and previous accesses which is available to model for user behavior analysis.
\end{itemize}
\section{Experimental Evaluation and Discussion}\label{secres}
\subsection{Experimental Setup}\label{subsecresa}
A series of experiments are carried out over a server machine comprising two Intel\textsuperscript{\textregistered} Xeon\textsuperscript{\textregistered} Silver 4114 CPU encompassing 40-core processor employing clock speed of 2.20 GHz. This computational system utilizes a 64-bit version of Ubuntu 18.04 LTS OS. The system is equipped with 128GB of main memory RAM. The software environment employed for the execution of experimental work is Python 3.6.8. For research purposes, a glimpse of the performance of the proposed federated model is showcased using diverse machine learning-based deep learning approaches such as ANN, CNN, and DNN using ReLu activation, Adam optimizer, and a final softmax output layer. The model is trained and tested over an 80:20 ratio of the dataset for analysis purposes. The enactment of the proposed strategy is tallied by focusing on two sorts of measures: the accuracy acquired after a fixed number of iterations (50) and the epochs (90) for two separate scenarios comprising $u_k =5$ users and $u_k =10$ users, respectively. A track of accuracy value and loss value after each round of the global model is also maintained.
\subsection{Dataset}\label{subsecresb}
To assess the model performance, an extended CMU CERT synthetic insider threat dataset r4.2 \cite{dataset} is utilized. The dataset is combating 10,000 training samples where each instance is comprised of a total of 13 current and historical features of every user as shown in Table \ref{dataset}.
\begin{table}[!htbp]
\centering
    \caption{Dataset description}\label{dataset}%
    \resizebox{1.0\textwidth}{!}{
        \begin{tabular}{|c|c|c|c|c|c|c|c|c|c|c|c|c|}
			\hline
          Profession & No. of & Type of & $DL$ & $HD$ & $LR$ & How & How & $DR$ & $LRa$ & $UT$ & $LC$ & Class\\
              & \# request & request &  &  &  & many LD & Frequently &  &  &  &  & \\
		    \hline 
            \hline
             5 & 162 & 4 & 15 & 1 & 1 & 9 & 0 & 12 & 6.0 & 2 & 5 & 1 \\ \hline
             2 & 18 & 2 & 35 & 1 & 1 & 4 & 1 & 1 & 22.0 & 2 & 3 & 1 \\ \hline
             2 & 94 & 1 & 10 & 0 & 0 & 7 & 18 & 10 & 7.0 & 2 & 4 & 1 \\ \hline
             5 & 110 & 1 & 10 & 1 & 1 & 4 & 18 & 5 & 4.0 & 1 & 3 & 1 \\ \hline
            \vdots & \vdots & \vdots & \vdots & \vdots & \vdots & \vdots & \vdots & \vdots & \vdots & \vdots & \vdots & \vdots \\ \hline
            5 & 302 & 5 & 30 & 0 & 0 & 2 & 16 & 8 & 1.0 & 0 & 1 & 2 \\ \hline
				\noalign{\smallskip}
    	   \end{tabular}}
    \footnotesize{$DL$: Data Limit; $HD$: Historical Data; \textit{$LR$}: Leaked Records; \textit{$DR$}: Data Retention; \textit{$LRa$}: Leak Ratio; \textit{$UT$}: User Type; \textit{$LC$}: Leak Channel;}
\end{table}
\par Current details describe \textit{profession, number of requests, type of requests, data limit, and user type} whereas historical features are \textit{historical data, leakage record, how many times data leaked, how frequently asking data, data retention, leak ratio, and leak channel}. Prominently, users can be classified into three classes which are: \textit{‘malicious’, ‘non-malicious’, and ‘unknown.’}
\subsection{Performance Parameters Analysis}\label{subsecresc}
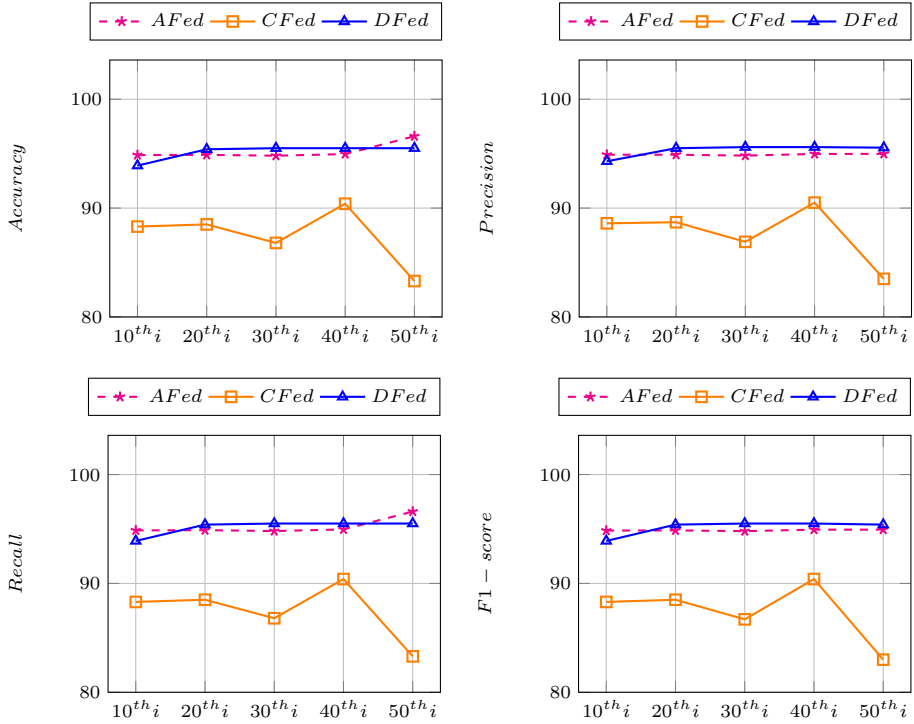
\begin{figure*}[!htbp]
\scriptsize
\subfigure{
\begin{tikzpicture}
\begin{axis}[
 ymin=80,
ymax = 100,
enlarge y limits={upper,value=0.18},
symbolic x coords={$10^{th}i$,$20^{th}i$,$30^{th}i$,$40^{th}i$,$50^{th}i$},
xtick=data,
height=5.0cm,
width=6.0cm,
grid=major,
ylabel={$Accuracy$},
legend style={at={(1.0,1.15)},
      anchor=east,legend columns=3}
]

\addplot+[mark=star, color=magenta,style=dashed,mark options={fill=white},draw=magenta, thick]  coordinates {
($10^{th}i$, 94.87) ($20^{th}i$, 94.89)  ($30^{th}i$, 94.81)  ($40^{th}i$, 94.96)  ($50^{th}i$, 96.60)
};
\addplot+[mark=square, color=orange,mark options={fill=white},draw=orange, thick] coordinates {
($10^{th}i$, 88.30) ($20^{th}i$, 88.50)  ($30^{th}i$, 86.80)  ($40^{th}i$, 90.40)  ($50^{th}i$, 83.30)
};
\addplot+[mark=triangle, color=blue,mark options={fill=white},draw=blue, thick]  coordinates {
($10^{th}i$, 93.90) ($20^{th}i$, 95.40)  ($30^{th}i$, 95.50)  ($40^{th}i$, 95.50)  ($50^{th}i$, 95.50)
};
\legend{$AFed$,$CFed$,$DFed$}
\end{axis}
\end{tikzpicture}
}
\hspace{0.05cm}
\subfigure{
\begin{tikzpicture}
\begin{axis}[
 ymin=80,
ymax = 100,
enlarge y limits={upper,value=0.18},
symbolic x coords={$10^{th}i$,$20^{th}i$,$30^{th}i$,$40^{th}i$,$50^{th}i$},
xtick=data,
height=5.0cm,width=6.0cm,
grid=major,
ylabel={$Precision$},
legend style={at={(1.0,1.15)},
      anchor=east,legend columns=3}
]

\addplot+[mark=star, color=magenta,style=dashed,mark options={fill=white},draw=magenta, thick] coordinates {
($10^{th}i$, 94.90) ($20^{th}i$, 94.90)  ($30^{th}i$, 94.82)  ($40^{th}i$, 94.97)  ($50^{th}i$, 94.98)
};
\addplot+[mark=square, color=orange,mark options={fill=white},draw=orange, thick] coordinates {
($10^{th}i$, 88.60) ($20^{th}i$, 88.70)  ($30^{th}i$, 86.90)  ($40^{th}i$, 90.50)  ($50^{th}i$, 83.50)
};

\addplot+[mark=triangle, color=blue,mark options={fill=white},draw=blue, thick] coordinates {
($10^{th}i$, 94.30) ($20^{th}i$, 95.50)  ($30^{th}i$, 95.60)  ($40^{th}i$, 95.60)  ($50^{th}i$, 95.55)
};

\legend{$AFed$,$CFed$,$DFed$}
\end{axis}
\end{tikzpicture}
}
\\
\subfigure{
\begin{tikzpicture}
\begin{axis}[
 ymin=80,
ymax = 100,
enlarge y limits={upper, value=0.18},
symbolic x coords={$10^{th}i$,$20^{th}i$,$30^{th}i$,$40^{th}i$,$50^{th}i$},
xtick=data,
height=5.0cm,width=6.0cm,
grid=major,
ylabel={$Recall$},
legend style={at={(1.0,1.15)},
      anchor=east,legend columns=3}
]

\addplot+[mark=star, color=magenta, style=dashed, mark options={fill=white},draw=magenta, thick]  coordinates {
($10^{th}i$, 94.87) ($20^{th}i$, 94.89)  ($30^{th}i$, 94.81)  ($40^{th}i$, 94.96)  ($50^{th}i$, 96.60)
};
\addplot+[mark=square, color=orange,mark options={fill=white},draw=orange, thick] coordinates {
($10^{th}i$, 88.30) ($20^{th}i$, 88.50)  ($30^{th}i$, 86.80)  ($40^{th}i$, 90.40)  ($50^{th}i$, 83.30)
};
\addplot+[mark=triangle, color=blue,mark options={fill=white},draw=blue, thick]  coordinates {
($10^{th}i$, 93.90) ($20^{th}i$, 95.40)  ($30^{th}i$, 95.50)  ($40^{th}i$, 95.50)  ($50^{th}i$, 95.50)
};
\legend{$AFed$,$CFed$,$DFed$}
\end{axis}
\end{tikzpicture}
}
\hspace{0.05cm}
\subfigure{
\begin{tikzpicture}
\begin{axis}[
 ymin=80,
ymax = 100,
enlarge y limits={upper,value=0.18},
symbolic x coords={$10^{th}i$,$20^{th}i$,$30^{th}i$,$40^{th}i$,$50^{th}i$},
xtick=data,
height=5.0cm,width=6.0cm,
grid=major,
ylabel={$F1-score$},
legend style={at={(1.0,1.15)},
      anchor=east,legend columns=3}
]

\addplot+[mark=star, color=magenta,style=dashed, mark options={fill=white},draw=magenta, thick] coordinates {
($10^{th}i$, 94.85) ($20^{th}i$, 94.87)  ($30^{th}i$, 94.80)  ($40^{th}i$, 94.94)  ($50^{th}i$, 94.94)
};
\addplot+[mark=square, color=orange,mark options={fill=white},draw=orange, thick] coordinates {
($10^{th}i$, 88.30) ($20^{th}i$, 88.50)  ($30^{th}i$, 86.70)  ($40^{th}i$, 90.40)  ($50^{th}i$, 83.00)
};

\addplot+[mark=triangle, color=blue,mark options={fill=white},draw=blue, thick] coordinates {
($10^{th}i$, 93.90) ($20^{th}i$, 95.40)  ($30^{th}i$, 95.50)  ($40^{th}i$, 95.50)  ($50^{th}i$, 95.40)
};

\legend{$AFed$,$CFed$,$DFed$}
\end{axis}
\end{tikzpicture}
}
\caption{Comparison of various performance parameters over ($T =$ 50) iterations for $u_k =10$ users}
\label{figiteration}
\end{figure*}

Figure \ref{figiteration} presents the performance evaluation of FedMUP for different performance parameters Accuracy, Recall, Precision, and F1-score. It shows a comparative analysis of all performance parameters considering five different points of evaluation at ($10^{th}$, $20^{th}$, $30^{th}$, $40^{th}$, and $50^{th}$) iteration, considering $u_k=10$ users. Model is enacted for different deep learning-based federated learning approaches which are AFed, CFed, and DFed. 
\par It highlights that the AFed is performing the assigned malicious user identification task with the highest efficiency and minimum computational requirement in comparison to other approaches. Moreover, the proposed FedMUP model is performing identification tasks proactively rather than after the occurrence of a data leakage event. Hence, the FedMUP model stands ahead of other methods to ensure data security by means of data allocation. Therefore, it stands fit for an efficient approach towards data utility, efficiency, identification of malicious commodities, and hence controlling the data leakages.
\par Table \ref{tab10user} and Table \ref{tab5user} are showcasing the performance parameters of the proposed FedMUP model using three different deep learning models; AFed, CFed, and DFed up-to $T = 50$ iterations, for $u_k =10$ and $u_k =5$ users, respectively. The observation noted is that, the AFed is outperforming over other defined methods for the dataset considered in this research.
\begin{table}[!htbp]
\centering
    \caption{Performance Metrics comparison of AFed, CFed and DFed for $u_k =10$ upto $T = 50$ iterations}\label{tab10user}%
    \resizebox{1.0\textwidth}{!}{
        \begin{tabular}{|c|c|c|c|c|c|c|c|c|c|c|c|c|}
			\hline
             \multirow{2}{*}{Iteration ($T$)} &  \multicolumn{4}{c}{AFed} &  \multicolumn{4}{|c|}{CFed} & \multicolumn{4}{|c|}{DFed}  \\ \cline{2-5} \cline{6-9} \cline{10-13} & Acc & Prec & Rec & F1 & Acc & Prec & Rec & F1 & Acc & Prec & Rec & F1\\
		    \hline 
            \hline
$0^{th}$ & 0.950 & 0.950 & 0.950 & 0.950 & 0.551 & 0.629 & 0.551 & 0.521 & 0.894 & 0.894 & 0.894 & 0.894 \\ \hline
$1^{st}$ & 0.950 & 0.950 & 0.950 & 0.949 & 0.604 & 0.757 & 0.604 & 0.581 & 0.904 & 0.906 & 0.904 & 0.904 \\ \hline
$2^{nd}$ & 0.949 & 0.950 & 0.949 & 0.949 & 0.744 & 0.780 & 0.744 & 0.746 & 0.904 & 0.905 & 0.904 & 0.904 \\ \hline
$3^{rd}$ & 0.950 & 0.951 & 0.950 & 0.950 & 0.836 & 0.838 & 0.836 & 0.836 & 0.898 & 0.897 & 0.898 & 0.897 \\ \hline
$4^{th}$ & 0.951 & 0.952 & 0.951 & 0.951 & 0.869 & 0.870 & 0.869 & 0.869 & 0.912 & 0.912 & 0.912 & 0.911 \\ \hline
$5^{th}$ & 0.950 & 0.951 & 0.950 & 0.949 & 0.813 & 0.813 & 0.813 & 0.813 & 0.927 & 0.929 & 0.927 & 0.926 \\ \hline
$6^{th}$ & 0.948 & 0.947 & 0.947 & 0.947 & 0.853 & 0.854 & 0.853 & 0.853 & 0.936 & 0.940 & 0.936 & 0.936 \\ \hline
$7^{th}$ & 0.948 & 0.948 & 0.948 & 0.948 & 0.862 & 0.862 &0.862 & 0.861 & 0.925 & 0.925 & 0.925 & 0.924 \\ \hline
$8^{th}$ & 0.950 & 0.951 & 0.950 & 0.950 & 0.845 & 0.850 & 0.845 & 0.842 & 0.936 & 0.940 & 0.936 & 0.936 \\ \hline
$9^{th}$ & 0.949 & 0.949 & 0.948 & 0.948 & 0.883 & 0.886 & 0.883 & 0.883 & 0.939 & 0.943 & 0.939 & 0.939 \\ \hline
$10^{th}$ & 0.949 & 0.950 & 0.949 & 0.949 & 0.875 & 0.875 & 0.875 & 0.875 & 0.946 & 0.948 & 0.946 & 0.945 \\ \hline
$11^{th}$ & 0.950 & 0.951 & 0.950 & 0.950 & 0.860 & 0.862 & 0.860 & 0.859 & 0.947 & 0.949 & 0.947 & 0.946 \\ \hline
$12^{th}$ & 0.949 & 0.950 & 0.949 & 0.949 & 0.746 & 0.783 & 0.746 & 0.721 & 0.946 & 0.948 & 0.946 & 0.945 \\ \hline
$13^{th}$ & 0.950 & 0.950 & 0.950 & 0.949 & 0.882 & 0.883 & 0.882 & 0.882 & 0.940 & 0.945 & 0.940 & 0.940 \\ \hline
$14^{th}$ & 0.949 & 0.949 & 0.949 & 0.949 & 0.812 & 0.821 & 0.812 & 0.805 & 0.948 & 0.950 & 0.948 & 0.948 \\ \hline
$15^{th}$ & 0.950 & 0.950 & 0.950 & 0.949 & 0.869 & 0.871 & 0.869 & 0.868 & 0.950 & 0.953 & 0.950 & 0.950 \\ \hline
$16^{th}$ & 0.949 & 0.949 & 0.949 & 0.949 & 0.766 & 0.779 & 0.766 & 0.755 & 0.950 & 0.951 & 0.950 & 0.949 \\ \hline
$17^{th}$ & 0.949 & 0.949 & 0.948 & 0.948 & 0.873 & 0.874 & 0.873 & 0.873 & 0.951 & 0.951 & 0.951 & 0.950 \\ \hline
$18^{th}$ & 0.948 & 0.948 & 0.948 & 0.948 & 0.881 & 0.882 & 0.881 & 0.880 & 0.949 & 0.950 & 0.949 & 0.949 \\ \hline
$19^{th}$ & 0.948 & 0.949 & 0.948 & 0.948 & 0.885 & 0.887 & 0.885 & 0.885 & 0.954 & 0.955 & 0.954 & 0.954 \\ \hline
$20^{th}$ & 0.948 & 0.949 & 0.948 & 0.948 & 0.891 & 0.892 & 0.891 & 0.891 & 0.954 & 0.954 & 0.954 & 0.953 \\ \hline
$21^{st}$ & 0.948 & 0.949 & 0.949 & 0.947 & 0.848 & 0.851 & 0.848 & 0.846 & 0.954 & 0.955 & 0.954 & 0.953 \\ \hline
$22^{nd}$ & 0.949 & 0.949 & 0.949 & 0.949 & 0.830 & 0.835 & 0.830 & 0.829 & 0.958 & 0.960 & 0.958 & 0.958 \\ \hline
$23^{rd}$ & 0.949 & 0.949 & 0.949 & 0.949 & 0.744 & 0.772 & 0.744 & 0.748 & 0.957 & 0.958 & 0.957 & 0.956 \\ \hline
$24^{th}$ & 0.946 & 0.945 & 0.946 & 0.945 & 0.901 & 0.902 & 0.901 & 0.901 & 0.955 & 0.956 & 0.955 & 0.955 \\ \hline
$25^{th}$ & 0.950 & 0.950 & 0.950 & 0.949 & 0.760 & 0.769 & 0.760 & 0.744 & 0.955 & 0.957 & 0.955 & 0.955 \\ \hline
$26^{th}$ & 0.949 & 0.950 & 0.949 & 0.949 & 0.853 & 0.857 & 0.853 & 0.851 & 0.956 & 0.957 & 0.956 & 0.956 \\ \hline
$27^{th}$ & 0.951 & 0.951 & 0.951 & 0.950 & 0.844 & 0.847 & 0.844 & 0.841 & 0.954 & 0.955 & 0.954 & 0.954 \\ \hline
$28^{th}$ & 0.951 & 0.951 & 0.951 & 0.950 & 0.831 & 0.835 & 0.831 & 0.828 & 0.955 & 0.956 & 0.955 & 0.955 \\ \hline
$29^{th}$ & 0.948 & 0.948 & 0.948 & 0.948 & 0.868 & 0.869 & 0.868 & 0.867 & 0.955 & 0.956 & 0.955 & 0.955 \\ \hline
$30^{th}$ & 0.949 & 0.949 & 0.949 & 0.949 & 0.885 & 0.886 & 0.885 & 0.885 & 0.956 & 0.956 & 0.956 & 0.955 \\ \hline
$31^{st}$ & 0.948 & 0.948 & 0.948 & 0.947 & 0.905 & 0.906 & 0.905 & 0.904 & 0.955 & 0.956 & 0.955 & 0.955 \\ \hline
$32^{nd}$ & 0.950 & 0.950 & 0.950 & 0.949 & 0.837 & 0.838 & 0.837 & 0.838 & 0.958 & 0.959 & 0.958 & 0.957 \\ \hline
$33^{rd}$ & 0.950 & 0.951 & 0.950 & 0.950 & 0.803 & 0.809 & 0.803 & 0.798 & 0.955 & 0.956 & 0.955 & 0.955 \\ \hline
$34^{th}$ & 0.950 & 0.951 & 0.950 & 0.950 & 0.838 & 0.843 & 0.838 & 0.835 & 0.957 & 0.958 & 0.957 & 0.956 \\ \hline
$35^{th}$ & 0.949 & 0.950 & 0.949 & 0.949 & 0.903 & 0.906 & 0.903 & 0.903 & 0.956 & 0.957 & 0.956 & 0.956 \\ \hline
$36^{th}$ & 0.951 & 0.951 & 0.951 & 0.951 & 0.821 & 0.825 & 0.821 & 0.817 & 0.957 & 0.958 & 0.957 & 0.956 \\ \hline
$37^{th}$ & 0.949 & 0.949 & 0.949 & 0.949 & 0.900 & 0.902 & 0.900 & 0.900 & 0.955 & 0.956 & 0.955 & 0.955 \\ \hline
$38^{th}$ & 0.948 & 0.949 & 0.948 & 0.948 & 0.881 & 0.884 & 0.881 & 0.880 & 0.956 & 0.957 & 0.956 & 0.955 \\ \hline
$39^{th}$ & 0.949 & 0.949 & 0.949 & 0.949 & 0.904 & 0.905 & 0.904 & 0.904 & 0.955 & 0.956 & 0.955 & 0.955 \\ \hline
$40^{th}$ & 0.951 & 0.951 & 0.951 & 0.951 & 0.849 & 0.853 & 0.849 & 0.846 & 0.954 & 0.954 & 0.954 & 0.953 \\ \hline
$41^{st}$ & 0.949 & 0.950 & 0.949 & 0.949 & 0.907 & 0.907 & 0.907 & 0.906 & 0.957 & 0.958 & 0.957 & 0.956 \\ \hline
$42^{nd}$ & 0.950 & 0.950 & 0.950 & 0.949 & 0.876 & 0.880 & 0.876 & 0.876 & 0.955 & 0.955 & 0.955 & 0.954 \\ \hline
$43^{rd}$ & 0.949 & 0.949 & 0.949 & 0.949 & 0.834 & 0.839 & 0.834 & 0.832 & 0.956 & 0.957 & 0.956 & 0.956 \\ \hline
$44^{th}$ & 0.950 & 0.951 & 0.950 & 0.950 & 0.874 & 0.874 & 0.874 & 0.873 & 0.952 & 0.952 & 0.952 & 0.952 \\ \hline
$45^{th}$ & 0.951 & 0.951 & 0.951 & 0.951 & 0.855 & 0.859 & 0.855 & 0.853 & 0.954 & 0.954 & 0.954 & 0.953 \\ \hline
$46^{th}$ & 0.950 & 0.951 & 0.950 & 0.950 & 0.823 & 0.826 & 0.823 & 0.820 & 0.954 & 0.954 & 0.954 & 0.953 \\ \hline
$47^{th}$ & 0.9501 & 0.950 & 0.951 & 0.950 & 0.844 & 0.847 & 0.844 & 0.841 & 0.955 & 0.956 & 0.955 & 0.955 \\ \hline
$48^{th}$ & 0.950 & 0.951 & 0.951 & 0.951 & 0.870 & 0.873 & 0.870 & 0.868 & 0.955 & 0.956 & 0.955 & 0.955 \\ \hline
$49^{th}$ & 0.966 & 0.949 & 0.966 & 0.949 & 0.833 & 0.835 & 0.833 & 0.830 & 0.955 & 0.955 & 0.955 & 0.954 \\ \hline
				\noalign{\smallskip}
	   \end{tabular}}
\end{table}
\begin{table}[!htbp]
\centering
    \caption{Performance Metrics comparison of AFed, CFed and DFed for $u_k =5$ upto $T = 50$ iterations}\label{tab5user}%
    \resizebox{1.0\textwidth}{!}{
        \begin{tabular}{|c|c|c|c|c|c|c|c|c|c|c|c|c|}
			\hline
             \multirow{2}{*}{Iteration ($T$)} &  \multicolumn{4}{c}{AFed} &  \multicolumn{4}{|c|}{CFed} & \multicolumn{4}{|c|}{DFed}  \\ \cline{2-5} \cline{6-9} \cline{10-13} & Acc & Prec & Rec & F1 & Acc & Prec & Rec & F1 & Acc & Prec & Rec & F1\\
		    \hline 
            \hline
$0^{th}$  & 0.947 & 0.947 & 0.948 & 0.947  & 0.387 & 0.498 & 0.387 & 0.227  & 0.906 & 0.906 & 0.906 & 0.905  \\ \hline
$1^{st}$  & 0.950 & 0.951 & 0.950 & 0.950  & 0.742 & 0.765 & 0.742 & 0.721  & 0.924 & 0.926 & 0.924 & 0.924  \\ \hline
$2^{nd}$  & 0.949 & 0.950 & 0.949 & 0.949  & 0.742 & 0.757 & 0.742 & 0.737  & 0.923 & 0.924 & 0.923 & 0.923  \\ \hline
$3^{rd}$  & 0.949 & 0.949 & 0.949 & 0.949  & 0.559 & 0.650 & 0.559 & 0.516  & 0.928 & 0.930 & 0.928 & 0.928  \\ \hline
$4^{th}$  & 0.948 & 0.948 & 0.948 & 0.948  & 0.804 & 0.809 & 0.804 & 0.800  & 0.928 & 0.930 & 0.928 & 0.928  \\ \hline
$5^{th}$  & 0.948 & 0.949 & 0.948 & 0.948  & 0.856 & 0.857 & 0.856 & 0.856  & 0.930 & 0.932 & 0.930 & 0.930  \\ \hline
$6^{th}$  & 0.948 & 0.949 & 0.948 & 0.948  & 0.641 & 0.684 & 0.641 & 0.624  & 0.929 & 0.931 & 0.929 & 0.929  \\ \hline
$7^{th}$  & 0.940 & 0.941 & 0.940 & 0.939  & 0.840 & 0.846 & 0.840 & 0.841  & 0.931 & 0.932 & 0.931 & 0.930  \\ \hline
$8^{th}$  & 0.949 & 0.950 & 0.949 & 0.949  & 0.838 & 0.841 & 0.838 & 0.836  & 0.931 & 0.932 & 0.931 & 0.930  \\ \hline
$9^{th}$  & 0.948 & 0.950 & 0.948 & 0.948  & 0.824 & 0.827 & 0.824 & 0.821  & 0.930 & 0.932 & 0.930 & 0.930  \\ \hline
$10^{th}$  & 0.946 & 0.946 & 0.946 & 0.946  & 0.851 & 0.852 & 0.851 & 0.850  & 0.930 & 0.933 & 0.930 & 0.929  \\ \hline
$11^{th}$  & 0.949 & 0.949 & 0.949 & 0.948  & 0.822 & 0.828 & 0.822 & 0.820  & 0.927 & 0.929 & 0.927 & 0.927  \\ \hline
$12^{th}$  & 0.947 & 0.947 & 0.947 & 0.947  & 0.864 & 0.864 & 0.864 & 0.864  & 0.930 & 0.932 & 0.930 & 0.930  \\ \hline
$13^{th}$  & 0.949 & 0.950 & 0.949 & 0.949  & 0.842 & 0.845 & 0.842 & 0.841  & 0.922 & 0.925 & 0.922 & 0.923  \\ \hline
$14^{th}$  & 0.946 & 0.946 & 0.946 & 0.946  & 0.811 & 0.817 & 0.811 & 0.807  & 0.928 & 0.930 & 0.928 & 0.928  \\ \hline
$15^{th}$  & 0.948 & 0.948 & 0.948 & 0.947  & 0.865 & 0.866 & 0.865 & 0.865  & 0.930 & 0.940 & 0.930 & 0.930  \\ \hline
$16^{th}$  & 0.946 & 0.946 & 0.946 & 0.946  & 0.826 & 0.833 & 0.826 & 0.824  & 0.926 & 0.929 & 0.926 & 0.926  \\ \hline
$17^{th}$  & 0.949 & 0.949 & 0.949 & 0.948  & 0.735 & 0.750 & 0.735 & 0.714  & 0.929 & 0.931 & 0.929 & 0.929  \\ \hline
$18^{th}$  & 0.949 & 0.949 & 0.949 & 0.949  & 0.807 & 0.813 & 0.807 & 0.804  & 0.925 & 0.926 & 0.925 & 0.925  \\ \hline
$19^{th}$  & 0.945 & 0.945 & 0.945 & 0.945  & 0.863 & 0.866 & 0.863 & 0.863  & 0.928 & 0.933 & 0.928 & 0.928  \\ \hline
$20^{th}$  & 0.946 & 0.946 & 0.946 & 0.946  & 0.826 & 0.828 & 0.826 & 0.825  & 0.928 & 0.934 & 0.928 & 0.928  \\ \hline
$21^{st}$  & 0.946 & 0.946 & 0.946 & 0.946  & 0.831 & 0.833 & 0.831 & 0.830  & 0.930 & 0.940 & 0.930 & 0.931  \\ \hline
$22^{nd}$  & 0.948 & 0.949 & 0.948 & 0.948  & 0.852 & 0.854 & 0.852 & 0.851  & 0.924 & 0.926 & 0.924 & 0.924  \\ \hline
$23^{rd}$  & 0.947 & 0.947 & 0.947 & 0.946  & 0.778 & 0.783 & 0.778 & 0.771  & 0.930 & 0.937 & 0.930 & 0.930  \\ \hline
$24^{th}$  & 0.948 & 0.948 & 0.948 & 0.948  & 0.863 & 0.864 & 0.863 & 0.863  & 0.926 & 0.930 & 0.926 & 0.926  \\ \hline
$25^{th}$  & 0.947 & 0.947 & 0.947 & 0.947  & 0.866 & 0.868 & 0.866 & 0.865  & 0.928 & 0.933 & 0.928 & 0.928  \\ \hline
$26^{th}$  & 0.945 & 0.946 & 0.945 & 0.945  & 0.860 & 0.861 & 0.860 & 0.860  & 0.925 & 0.927 & 0.925 & 0.925  \\ \hline
$27^{th}$  & 0.947 & 0.947 & 0.947 & 0.947  & 0.818 & 0.821 & 0.818 & 0.816  & 0.928 & 0.931 & 0.928 & 0.928  \\ \hline
$28^{th}$  & 0.945 & 0.945 & 0.945 & 0.945  & 0.842 & 0.847 & 0.842 & 0.843  & 0.924 & 0.927 & 0.924 & 0.925  \\ \hline
$29^{th}$  & 0.947 & 0.947 & 0.947 & 0.947  & 0.848 & 0.850 & 0.848 & 0.847  & 0.928 & 0.931 & 0.928 & 0.928  \\ \hline
$30^{th}$  & 0.948 & 0.948 & 0.948 & 0.948  & 0.860 & 0.860 & 0.860 & 0.860  & 0.929 & 0.934 & 0.929 & 0.929  \\ \hline
$31^{st}$  & 0.948 & 0.948 & 0.948 & 0.948  & 0.859 & 0.861 & 0.859 & 0.859  & 0.928 & 0.933 & 0.928 & 0.928  \\ \hline
$32^{nd}$  & 0.949 & 0.950 & 0.949 & 0.949  & 0.855 & 0.857 & 0.855 & 0.855  & 0.930 & 0.937 & 0.930 & 0.930  \\ \hline
$33^{trd}$  & 0.946 & 0.946 & 0.946 & 0.946  & 0.854 & 0.856 & 0.854 & 0.854  & 0.927 & 0.932 & 0.927 & 0.927  \\ \hline
$34^{th}$  & 0.944 & 0.944 & 0.944 & 0.944  & 0.849 & 0.853 & 0.849 & 0.849  & 0.928 & 0.932 & 0.928 & 0.928  \\ \hline
$35^{th}$  & 0.948 & 0.948 & 0.948 & 0.948  & 0.870 & 0.873 & 0.870 & 0.869  & 0.927 & 0.931 & 0.927 & 0.927  \\ \hline
$36^{th}$  & 0.948 & 0.948 & 0.948 & 0.947  & 0.867 & 0.870 & 0.867 & 0.867  & 0.928 & 0.931 & 0.928 & 0.928  \\ \hline
$37^{th}$  & 0.948 & 0.948 & 0.948 & 0.948  & 0.772 & 0.772 & 0.772 & 0.771  & 0.926 & 0.930 & 0.926 & 0.926  \\ \hline
$38^{th}$  & 0.948 & 0.948 & 0.948 & 0.948  & 0.824 & 0.829 & 0.824 & 0.823  & 0.927 & 0.931 & 0.927 & 0.927  \\ \hline
$39^{th}$  & 0.946 & 0.946 & 0.946 & 0.946  & 0.866 & 0.869 & 0.866 & 0.865  & 0.925 & 0.928 & 0.925 & 0.925  \\ \hline
$40^{th}$  & 0.949 & 0.949 & 0.949 & 0.948  & 0.802 & 0.806 & 0.802 & 0.801  & 0.926 & 0.929 & 0.926 & 0.926  \\ \hline
$41^{st}$  & 0.948 & 0.948 & 0.948 & 0.948  & 0.793 & 0.802 & 0.793 & 0.790  & 0.925 & 0.928 & 0.925 & 0.925  \\ \hline
$42^{nd}$  & 0.948 & 0.948 & 0.948 & 0.948  & 0.804 & 0.808 & 0.804 & 0.803  & 0.926 & 0.930 & 0.926 & 0.926  \\ \hline
$43^{rd}$  & 0.947 & 0.947 & 0.947 & 0.947  & 0.834 & 0.835 & 0.834 & 0.834  & 0.928 & 0.935 & 0.928 & 0.929  \\ \hline
$44^{th}$  & 0.949 & 0.949 & 0.949 & 0.949  & 0.800 & 0.804 & 0.800 & 0.798  & 0.925 & 0.927 & 0.925 & 0.926  \\ \hline
$45^{th}$  & 0.951 & 0.951 & 0.951 & 0.951  & 0.823 & 0.829 & 0.823 & 0.823  & 0.929 & 0.936 & 0.929 & 0.929  \\ \hline
$46^{th}$  & 0.945 & 0.945 & 0.945 & 0.945  & 0.815 & 0.831 & 0.815 & 0.817  & 0.929 & 0.937 & 0.929 & 0.930  \\ \hline
$47^{th}$  & 0.950 & 0.950 & 0.950 & 0.949  & 0.858 & 0.861 & 0.858 & 0.858  & 0.926 & 0.930 & 0.926 & 0.926  \\ \hline
$48^{th}$  & 0.947 & 0.947 & 0.947 & 0.947  & 0.813 & 0.821 & 0.813 & 0.811  & 0.926 & 0.931 & 0.926 & 0.926  \\ \hline
$49^{th}$  & 0.947 & 0.948 & 0.947 & 0.947  & 0.868 & 0.874 & 0.868 & 0.867  & 0.926 & 0.930 & 0.926 & 0.926  \\ \hline
				\noalign{\smallskip}
	   \end{tabular}}
\end{table}
\par
\begin{figure}[ht!]%
    \begin{center}
        \subfigure[Success rate v/s Epochs ($\xi$)] {
        \includegraphics[height=4cm, width=0.5\textwidth]{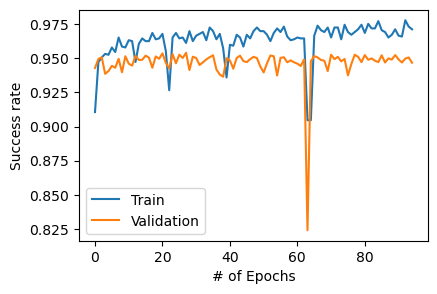}
        }%
        \subfigure[Loss v/s Epochs ($\xi$)]{
        \includegraphics[height=4cm, width=0.5\textwidth]{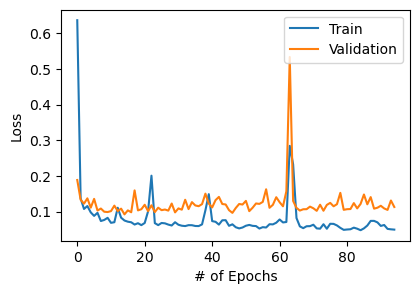}
        }%
    \end{center}       
\caption{Comparison of Success rate and Loss over Epochs ($\xi$) for $u_k = 5$} \label{acclossu5}
\end{figure}

\begin{figure}[ht!]%
    \begin{center}
        \subfigure[Success rate v/s Epochs ($\xi$)]{
        \includegraphics[height=4cm, width=0.5\textwidth]{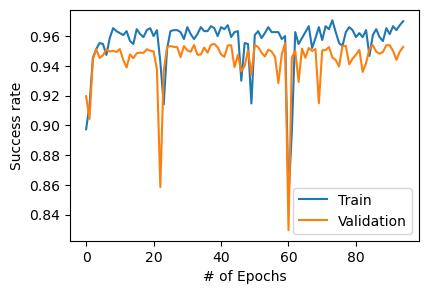}
        }%
        \subfigure[Loss v/s Epochs ($\xi$)]{
        \includegraphics[height=4cm, width=0.5\textwidth]{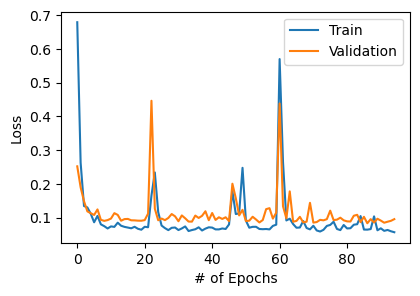}
        }%
    \end{center}       
\caption{Comparison of Success rate and Loss over Epochs ($\xi$) for $u_k =10$} \label{acclossu10}
\end{figure}

 In particular, Figure \ref{acclossu5} and Figure \ref{acclossu10} display the success rate and the expected loss in two separate model training and validation scenarios, holding $u_k =5$ and $u_k =10$ users, where the communication epochs ($\xi$) are set as $\xi$ = 90 to inspect the impact of proposed model interpretation, respectively in each scenario. In experiments, the whole dataset, i.e., CMU CERT r4.2, is diverged in such a way that each player gets an equal size of fragments.
\begin{table}[!htbp]
\small
\centering
    \caption{Performance metrics of global model for different users over $\xi =$ 90}\label{tabgm}%
    \resizebox{1.0\textwidth}{!}{
        \begin{tabular}{l|cccc|cccc}
			\hline \hline
             \multirow{2}{*}{Approach} & \multicolumn{4}{c}{$u_k =5$} &  \multicolumn{4}{|c}{$u_k =10$} \\ \cline{2-9} & Success & Loss & Time & Memory & Success & Loss & Time & Memory \\
             & rate & & &  & rate & & & \\
		    \hline 
            \hline
            AFed & 96.40  & 0.0682  & 313.135  & 15.79 & 96.73 & 0.0659 & 367.728 & 18.92 \\ \hline
            CFed & 91.30  & 0.0392  & 201.315  & 15.08 & 90.05 & 0.0470 & 377.125 & 23.76 \\ \hline
            DFed & 92.80  & 0.0386  & 288.811  & 9.28 & 95.10 & 0.0285 & 469.750 & 11.62 \\ \hline \hline
				\noalign{\smallskip}
	   \end{tabular}}
\end{table}
\par Table \ref{tabgm} up-brings the overall health of the proposed federated learning-oriented malicious user prediction model performance parameter success rate, loss value, time, and memory for $u_k =5$ users and $u_k =10$ users respectively. This table demonstrates the efficiency and functioning of the model without any over-fitting of user features in the model.
\subsection{Comparison}\label{subsecresd}
\par A comparison of the proposed FedMUP model is performed with already existing diverse state-of-the-art works such as \textit{Attribute/Behavior-Based Access Control} (ABBAC) scheme \cite{afshar2021incorporating}, \textit{Guilty Agent Model} (GAM) \cite{5487521}, \textit{Dynamic Threshold based Information Leaker Identification scheme} (DT-ILIS) \cite{gupta2019dynamic}, \textit{Machine Learning and Probabilistic Analysis Based Model} (MLPAM) \cite{gupta2020mlpam},  \textit{Quantum Machine Learning driven Malicious User Prediction Model} (QM-MUP) \cite{9865138}, \textit{A learning-oriented DLP system based on XGBoost classification} \cite{gupta_Kush_2020}, and an outsourced cloud-based secure communication model for advanced privacy preserving data computing and protection (SeCoM) \cite{SeCom10130499}. 
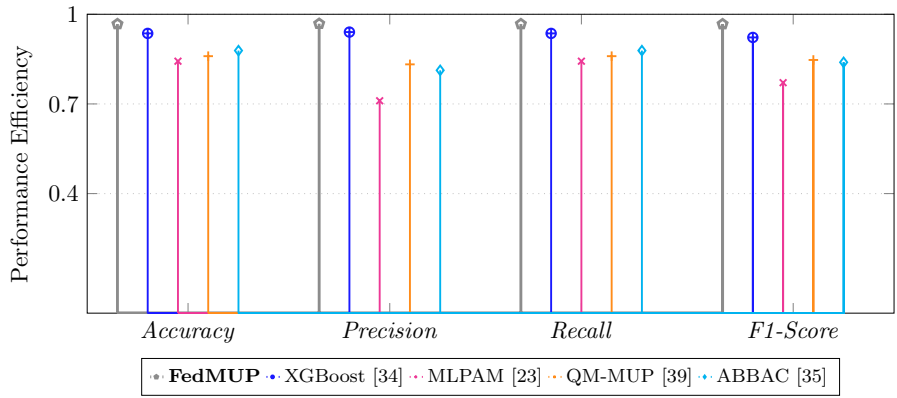
\begin{figure}[!htbp]
\begin{tikzpicture}[node distance = 1cm,auto,scale=.90, transform shape]
\pgfplotsset{every axis y label/.append style={rotate=180,yshift=10.5cm}}
\begin{axis}[
      axis on top=false,
        height=6.0cm,width=13.5cm,
      xmin=9, xmax=89,
      ymin=0, ymax=1,
      ytick={0.4,0.7,1},
      xtick={19,39,58,79},
      xticklabels={\textit{Accuracy}, \textit{Precision}, \textit{Recall}, \textit{F1-Score}},
        ycomb,
        ylabel near ticks, yticklabel pos=left,
      ylabel={Performance Efficiency},
       legend image post style={scale=0.3}, 
       legend style={at={(0.5,-0.15)},
      anchor=north,legend columns=5},
      ymajorgrids=true,
      grid style=dotted,
          ]
\addplot+[mark=pentagon, mark options={fill=green},fill=pink, draw=gray!90,  very thick] 
coordinates
{(12,.9673) (32,.9684) (52,.9673) (72,.9670) }
\closedcycle;
\addlegendentry{\footnotesize \textbf{FedMUP}}

\addplot+[mark=oplus,mark options={fill=blue},fill=blue,draw=blue!90, thick]
coordinates
 {(15,.9360) (35,.9404) (55,.9360) (75,.9225) }
\closedcycle;
\addlegendentry{\footnotesize XGBoost \cite{gupta_Kush_2020}}

\addplot+[mark=x,mark options={fill=pink},fill=pink,draw=magenta!90, thick] 
coordinates
 {(18,.8430) (38,.7106) (58,.8430) (78,.7711) }
\closedcycle;
\addlegendentry{\footnotesize MLPAM \cite{gupta2020mlpam}}

\addplot+[mark=+,mark options={fill=white},fill=red,draw=orange!90, thick] 
coordinates
 {(21,.8598) (41,.8324) (61,.8598) (81,.8471) }
\closedcycle;
\addlegendentry{\footnotesize QM-MUP \cite{9865138}}
\addplot+[mark=diamond,mark options={fill=cyan},fill=cyan,draw=cyan!90, thick] 
coordinates
 {(24,.8785) (44,.8125) (64,.8785) (84,.8393)}
\closedcycle;
\addlegendentry{\footnotesize ABBAC \cite{afshar2021incorporating}}
\end{axis}
\end{tikzpicture}
\caption{Performance Parameters: FedMUP v/s state-of-the-arts} \label{figcomp}
\end{figure}
\par In Figure \ref{figcomp} a brief comparison of various performance parameters \textit{accuracy}, \textit{precision}, \textit{recall}, and \textit{f1-score} of proposed FedMUP model is made with state-of-the-art works. Performance parameters are computed for all approaches considering all agents. Observations noted are that firstly, in the case of \textit{Accuracy}, the performance of the proposed FedMUP model is enhanced in the range 1.23$\%$ to 14.32$\%$ than state-of-the-art works. For \textit{Precision} improvement lies in the range 1.62$\%$ to 17.88$\%$. In the case of \textit{Recall} a hike of 1.5$\%$ to 14.32$\%$ is noted and for \textit{F1-Score} enhancement lies in the range of 1.58$\%$ to 18.35$\%$. It is evident from this figure that the performance of the FedMUP model is dominating over all state-of-the-art works. Therefore, the FedMUP model seems to serve best for proactive real-time identification of malicious users and data demands. It yields efficiency because of the computational efficiency of deep learning over existing traditional approaches for the same purpose.
\begin{table*}[!htbp] 
\small
	\centering
		\caption{Summarized Performance and Complexity}
		\label{taboverall}
        \resizebox{0.9\textwidth}{!}{    
			\begin{tabular}{l||cccc||c}
				\hline \hline
				Models & Accuracy & Precision & Recall & F1-Score & Complexity\\ 
			  \hline \hline
			 GAM \cite{5487521} & 59.00 &  55.00 & 96.00 & 70.00 & ${O}(zm + |\sum_{j=1}^{m}d_{j}|)$\\ \hline
			 DT-ILIS \cite{gupta2019dynamic} & 64.00 & 59.00 & 97.00 & 73.00 & ${O}(z + |\sum_{j=1}^{m}d_{j}|)$   \\ \hline
			MLPAM \cite{gupta2020mlpam} & 84.30 & 71.06 & 84.30 & 77.11 & ${O}(|\sum_{j=1}^{m}d_{j}|)$  \\\hline
			QM-MUP \cite{9865138} & 85.98 & 83.24 & 85.98 & 84.71 & 
			${O}$($tL{N}$ $N^{\ast})$\\ \hline
            XGBoost \cite{gupta_Kush_2020} & 93.60 & 94.04 & 93.60 & 92.25 & 
			${O}$($tmxyN)$ \\ \hline
            SeCoM \cite{SeCom10130499} & 94.20 & 94.23 & 93.62 & 94.20 &  
			${O}(|\sum_{j=1}^{q}d_{j}|)$ \\ \hline
             \textbf{FedMUP} & 96.73 & 96.84 & 96.73 & 96.70 & 
			${O}(ntL\xi{N}^{\ast})$
			\\ \hline \hline
				\noalign{\smallskip}
		    \end{tabular}}
\end{table*}
\par Table \ref{taboverall} showcases a comparison of the overall performance parameters and computational complexity of the proposed FedMUP model with other state-of-the-art models. It is evident from the table that the proposed work is taking the lead with the highest value of performance parameters and lesser complexity. Hence, it can be stated that the FedMUP model is performing malicious user identification to ensure data security for communication in the cloud platform.
\subsection{Feature Analysis}\label{subsecrese}
An elaborate feature analysis of the work under consideration is conducted with a few of the existing state-of-the-art approaches GAM \cite{5487521}, DT-ILIS \cite{gupta2019dynamic}, MLPAM \cite{gupta2020mlpam}, ABBAC \cite{afshar2021incorporating}, QM-MUP \cite{9865138}, XGBoost \cite{gupta_Kush_2020}, and (SeCoM) \cite{SeCom10130499} in Table \ref{tabfeature}. ABBAC established an access control approach by utilizing various user behavior details to find malicious entities. GAM employed the concept of data allocation with the principle of minimum overlapping to find the malicious entity based on the data allocation pattern. DT-ILIS proposed an approach to identify the data leaker depending on the preset threshold value. MLPAM presented a sophisticated guilty agent detection strategy using data distribution and probability metrics. QM-MUP incorporated an all-in-first-time approach to identify guilty ones proactively by deploying the computability of quantum mechanics in the form of qubits and quantum Pauli gates. XGBoost devised an approach for data classification as restricted, and non-restricted for secure communication. SeCoM improvised a privacy-preserving method for secure healthcare data protection in the cloud and IoT systems by minimizing the threat of data leakage, identifying, and terminating malicious entities against data leakage, and addressing security threats. 
\begin{table*}[!htbp] 
\scriptsize
\caption{FedMUP Features Analysis comparison with state-of-the-art works}
\label{tabfeature}
\centering
\resizebox{1.0\textwidth}{!}{
\begin{tabular}{l||cccccccccc}
\hline \hline
Models & \textit{$NE$} & \textit{$DO$} & \textit{$U$} & \textit{$DT$} & \textit{$SDS$} & \textit{$SDA$} & \textit{$SDD$} & \textit{$SC$} & \textit{$MEP$} &\textit{$LR$} \\ \hline \hline 

GAM \cite{5487521} & $\Box$ & $\Box$ & $\boxtimes$ & $\boxtimes$ & $\times$ & $\times$ & $\times$ & $\times$ & $\Box$ & $\times$ \\ \hline 

DT-ILIS \cite{gupta2019dynamic} & $\Box$ & $\Box$ & $\boxtimes$ & $\Box$ & $\times$ & $\times$ & $\times$ & $\times$ & $\Box$ & $\times$ \\ \hline 

MLPAM \cite{gupta2020mlpam} & $\boxtimes$ & $\boxtimes$ & $\boxtimes$ & $\Box$ & $\surd$ & $\surd$ & $\surd$ & $\times$ & $\Box$ & $\times$ \\ \hline 

ABBAC \cite{afshar2021incorporating} & $\Box$ & $\Box$ & $\boxtimes$ & $\boxtimes$ & $\times$ & $\times$ & $\surd$ & $\times$ & $\Box$ & $\times$ \\ \hline 

QM-MUP \cite{9865138} & $\boxtimes$ & $\boxtimes$ & $\boxtimes$ & $\boxtimes$ & $\surd$ & $\surd$ & $\surd$ & $\surd$ & $\boxtimes$ & $\times$ \\ \hline 

XGBoost \cite{gupta_Kush_2020} & $\boxtimes$ & $\boxtimes$ & $\boxtimes$ & $\boxtimes$ & $\surd$ & $\surd$ & $\surd$ & $\times$ & $\boxtimes$ & $\surd$ \\ \hline 

SeCoM\cite{SeCom10130499} & $\boxtimes$ & $\boxtimes$ & $\boxtimes$ & $\boxtimes$ & $\surd$ & $\surd$ & $\surd$ & $\times$ & $\boxtimes$ & $\times$ \\ \hline 

\textbf{FedMUP} & $\boxtimes$ & $\boxtimes$ & $\boxtimes$ & $\boxtimes$ & $\surd$ & $\surd$ & $\surd$ & $\surd$ & $\boxtimes$ & $\surd$ \\ \hline 
\noalign{\smallskip}
\end{tabular}
}
\footnotesize{$\Box$: Single; $\boxtimes$: Multiple; \textit{$NE$}: Non-entrusted Entity; \textit{$DO$}: Data Owners; \textit{$U$}: Users; \textit{$SDS$}: Secure Data Storage; \textit{$SDA$}: Secure Data Analysis; \textit{$SDD$}: Secure Data Distribution; \textit{$SC$}: Secure Communication; \textit{$MEP$}: Malicious  Entity Prediction; \textit{$LR$}: Learning Rate; } 
\end{table*}
\par The proposed FedMUP model computes the malicious entity identification task with much higher efficiency and high value of various performance parameters. However, it considers, all the participating entities such as Data owners, Users, Cloud service providers, etc. to be non-entrusted. Besides, this analysis also exhibits that the FedMUP model retains the most elevated efficiency for noticing malicious users proactively just before a data leakage incidence, unlikely to GAM \cite{5487521}, DT-ILIS \cite{gupta2019dynamic}, MLPAM \cite{gupta2020mlpam}, ABBAC \cite{afshar2021incorporating},  XGBoost \cite{gupta_Kush_2020}, and SeCoM \cite{SeCom10130499} which demarcate the malicious commodity after a data violation event. Moreover, the proposed FedMUP model is discharging the responsibility of data protection over cloud platforms with the most improved success rate in unveiling evil users. Thus, the FedMUP model stands superior in comparison to the other state-of-the-art works, in all means to ensure secure data allocation, distribution, sharing, and communication.
\section{Conclusion and Future Work}\label{seccon}
In this article, a novel FedMUP model is presented for malicious user prediction. This model employs the intense computational ability of federated machine learning to scrutinize user behavior. The proposed FedMUP model works proactively to ensure data security and data availability with ease without performance degradation through local modal training and sharing learned values, rather than actual raw data to build an enriched global model by averaging local updates, which ultimately reduces the need for actual sensitive data communication and therefore acquires enhanced data security. The critical performance parameters like accuracy, precision, recall, f1-score, and loss values are evaluated for multiple local epochs over numerous iterations to showcase the model's capability. The future goal is to seek extension for enhanced security while sharing the local models over cloud platforms and to develop an adaptive learning supervised privacy-preserving approach for data protection.

\bibliography{mybibfile}

\end{document}